\documentclass[%
 aip,
 amsmath,amssymb,
 reprint,%
]{revtex4-1}

\usepackage{graphicx}
\usepackage{dcolumn}
\usepackage{bm}

\usepackage[utf8]{inputenc}
\usepackage[T1]{fontenc}
\usepackage{mathptmx}
\usepackage{etoolbox}

\usepackage{tikz}
\usepackage{amsmath} 
\usetikzlibrary{shapes.geometric, positioning, arrows.meta, calc}
\usepackage{amsfonts} 

\definecolor{bannerblue}{HTML}{1E3A8A}
\definecolor{accentblue}{HTML}{3B82F6}
\definecolor{accentteal}{HTML}{0F766E}
\definecolor{accentpurple}{HTML}{6D28D9}
\definecolor{accentpink}{HTML}{BE185D}
\definecolor{boxbg}{HTML}{F8FAFC}
\definecolor{novelbg}{HTML}{F0FDF4}
\definecolor{novelgreen}{HTML}{15803D}
\definecolor{noveltext}{HTML}{14532D}
\definecolor{bordergray}{HTML}{CBD5E1}
\definecolor{textdark}{HTML}{1E293B}
\definecolor{textmuted}{HTML}{475569}

\makeatletter
\def\@email#1#2{%
 \endgroup
 \patchcmd{\titleblock@produce}
  {\frontmatter@RRAPformat}
  {\frontmatter@RRAPformat{\produce@RRAP{*#1\href{mailto:#2}{#2}}}\frontmatter@RRAPformat}
  {}{}
}%
\makeatother
\begin{document}

\preprint{AIP/123-QED}

\title[]{The impact of intraband carrier dynamics on the optical properties \\ of two-dimensional materials: I. General theory }

\author{A.O. Slobodeniuk}
\email{artur.slobodeniuk@matfyz.cuni.cz}
\author{T. Novotn\'{y}}
\email{tomas.novotny@matfyz.cuni.cz}
\affiliation{ 
Department of Condensed Matter Physics, Faculty of Mathematics and Physics,
Charles University, Ke Karlovu 5, CZ-121 16 Prague, Czech Republic}

\date{\today}

\begin{abstract}
We consider the generation of microscopic polarization in two-dimensional semiconductors under intense optical fields in the non-resonant regime. We demonstrate that the intraband motion of quasiparticles, driven by the electric field of the laser pulse, contributes substantially to the system's polarization. The effects of intraband motion are analyzed using the semiconductor Bloch equations. We propose a method for solving these equations by considering them in a reference frame co-moving along the direction of the electric field of the optical pulse. We demonstrate the developed technique on transition metal dichalcogenide monolayers irradiated by strong circularly-polarized infrared light. The solution is provided in the form of a power series expansion in a small parameter that naturally emerges in the system.  The implications of the results are examined, and the limitations of the approach are discussed. 
\end{abstract}

\maketitle

\section{Introduction}

The Semiconductor Bloch Equations (SBEs) serve as a fundamental theoretical framework for studying ultrafast nonlinear phenomena in solids. By describing the dynamics of electronic populations and coherences within the material’s band structure, SBEs enable detailed analysis of how intense and ultrashort optical pulses interact with semiconductors. This approach provides valuable insights into a wide range of nonlinear processes, such as high harmonic generation, nonlinear absorption, carrier excitation, and phase modulation, occurring on femtosecond timescales. The application of SBEs to these phenomena enhances our understanding of fundamental light-matter interactions in condensed matter systems and supports the development of advanced photonic and optoelectronic technologies, including ultrafast switches, frequency converters, and quantum information devices.

The SBE are a set of integro-differential equations for interband polarization and intraband carrier distributions of the electronic bands in semiconducting crystals. The SBE provide a versatile method that describes the dynamics of electron states in semiconductors \cite{Kira2006,Haug2009,Kira2011}, under the influence of terahertz (THz) fields \cite{Danielson2007,Golde2011,Vanska2015} and / or light pulses \cite{Cundiff1994,Ning1997,Smith2010,Malic2011}, which include the effects of electron-electron \cite{Bowden1995,Rochat2000,Booth2021} and electron-phonon interactions \cite{Kuznetsov1991,Forstener2003,Feldmann2009}. These equations are a powerful tool for exploring time-dependent processes in semiconductors in the subpicosecond time domain \cite{Kuznetsov1991_1,Kuznetsov1993,Jiang2009,Hohenleutner2015,Kalt2024,Koutensky2023}.  
The SBE have been developed to correctly describe the optical properties of semiconductors\cite{Haug1985,Schmitt-Rink1986,Muller1987,Schmitt-Rink1988} where the effects of Coulomb coupling between quasiparticles cannot be ignored. 
The optical Stark effect \cite{Zimmermann1990,Chemla1989}, the burning of spectral holes \cite{Nishimura1973,Paul1992,Meissner1993}, the four-wave mixing \cite{Lindberg1994,Lummer1996}, and other linear and nonlinear phenomena have been successfully explained with this approach. In these studies the interaction with light was considered in the approximation when only the inter-band coupling terms were taken into account. The intraband terms were omitted as non-dominant, due to i) the weak intensity of the light pulses, used in the experiments; ii) the near-resonant regime, when the energy of the incident photons is close to the band gap of the semiconductor; see a discussion in Ref.~[\onlinecite{Keldysh1994}].   

These conditions are violated in modern studies where the semiconductors are excited with high-power off-resonant THz and optical pulses, leading to nonlinear effects such as high harmonic generation \cite{Hohenleutner2015,Wu2015,Tancogne-Dejean2017,You2017,Peterka2023}, dynamical Bloch oscillations \cite{Shubert2014,McDonald2015,Liang2018}, and appearance of high-order sidebands \cite{Langer2016,Costello2021}. In these cases, both the intra- and the inter-band terms contribute significantly, which makes the analytical solutions of the corresponding SBE complicated. For such a type of problems, the solutions of the full SBE are performed only numerically \cite{Vampa2014,Luu2016,Schmidt2016,Katsch2020,Katsch2020_1,Lun2020,Meckbach2020,Kolesik2023}.  

Therefore, there is a lack of analytical solutions to the SBE in the regime of high-power off-resonant optical or THz pulses applied to semiconductor crystals, where the contribution of the intraband coupling term becomes important. To clarify the role of the intraband term, we derive the SBE in the most general form for the important case of two-dimensional semiconductors irradiated by an off-resonant strong optical pulse. We then customize these equations for a semiconducting transition-metal dichalcogenide (S-TMD) monolayer irradiated by a circularly polarized light pulse and analyze the role of intra- and inter-band coupling terms in them. We propose a method for accounting for the intraband term by considering the SBE in a moving coordinate frame in momentum space. Using this approach, we obtain an analytical expression for the polarization induced by the optical pump in the S-TMD monolayer. We observe that this result consists of two contributions, associated with the intra- and inter-band coupling terms in the SBE.
The impact of each term to the total result as a function of the pump-pulse parameters and material properties of the S-TMD monolayer is studied.  

The paper is organized as follows. In Sec.~\ref{sec:key_ideas} we provide a brief overview of new key ideas and methods discovered to solve the considered problem. In Sec.~\ref{sec:two_band_model} we introduce the Hamiltonian of a two-dimensional semiconductor in which the valence and conduction band electrons interact with i) the external electric field and ii) each other via the Coulomb interaction. After the Hamiltonian formulation of the problem we derive the SBE in the Hartree-Fock approximation in Sec.~\ref{sec:sbe}. We discuss the role of the material (Lagrangian) derivative appearing in these equations and analyze the general consequences of their solutions. 
In Sec.~\ref{sec:sbe_2d} we adapt the general SBE to the particular case of the S-TMD monolayer. 
Then, in Sec.~\ref{sec:pump} we consider the problem of irradiation of the S-TMD monolayer by an intensive off-resonant circularly-polarized optical pulse. Namely, we derive the equation for the interband polarization induced by the pulse, and develop a perturbative technique to solve it. The obtained results are  summarized in Sec.~\ref{sec:conclusions}. Finally, the technical details of the calculations of the main results of the study are shown in 3 Appendices.  

\section{Key ideas and scheme of solution}
\label{sec:key_ideas}

\vspace{-2mm}
The block diagram below illustrates the theoretical solver roadmap for the Semiconductor Bloch Equations.
\vspace{2mm}

\begin{tikzpicture}[
    basebox/.style={
        rectangle, 
        draw=bordergray, 
        fill=boxbg, 
        minimum width=8.2cm, 
        minimum height=1.4cm, 
        text width=7.7cm, 
        align=left, 
        rounded corners=4pt, 
        line width=0.6pt,
        inner sep=6pt
    },
    arrow/.style={
        ->, 
        >=Stealth, 
        line width=1.0pt, 
        color=textmuted
    }
]

    \node[basebox] (step1) at (0,0) {
        {\sffamily\bfseries\color{textdark}\small 1. Initial System: Semiconductor Bloch Equations (SBE) with intra- and interband terms, Eqs.~(\ref{eq:pcv}-\ref{eq:nv})}\\
        {\sffamily\footnotesize\color{textmuted} Tracks interband polarization $P_{cv}(\mathbf{k},t)$ and occupation numbers 
        of conduction $N_c(\mathbf{k},t)$ and valence $N_v(\mathbf{k},t)$ bands across 2D momentum space $\mathbf{k} = (k_x, k_y)$} ) 
    };
    \fill[accentblue] ($(step1.north west)+(0.01,-0.01)$) rectangle ($(step1.south west)+(0.12,0.01)$);

    \node[basebox, below=0.4cm of step1] (step2) {
        {\sffamily\bfseries\color{textdark}\small 2. Key Idea I: Material Derivative Mapping, Eq.~(\ref{eq:lagrangian_derivative})}\\
        {\sffamily\footnotesize\color{textmuted}Transforms the SBE using a Lagrangian description}\\
        {\sffamily\footnotesize\bfseries\color{accentblue!80!black}\hspace{2.2cm} $\mathcal{D} = \partial/\partial t + (e/\hbar)\mathbf{E}(t) \cdot \nabla_\mathbf{k}$}
    };
    \fill[accentblue!80!black] ($(step2.north west)+(0.01,-0.01)$) rectangle ($(step2.south west)+(0.12,0.01)$);

    \node[basebox, below=0.4cm of step2] (step3) {
        {\sffamily\bfseries\color{textdark}\small 3. System Reduction via Integral of Motion, Eq.~(\ref{eq:linearized_equation})}\\
        {\sffamily\footnotesize\color{textmuted}Identifies invariants to fully eliminate occupation variables, collapsing system into a single equation for $P_{cv}(\mathbf{k},t)$}
    };
    \fill[accentblue!50!black] ($(step3.north west)+(0.01,-0.01)$) rectangle ($(step3.south west)+(0.12,0.01)$);

    \node[basebox, below=0.4cm of step3] (step4) {
        {\sffamily\bfseries\color{textdark}\small 4. Key Idea II: Co-Moving Reference Frame, Eq.~(\ref{eq:unitary})}\\
        {\sffamily\footnotesize\color{textmuted}Transforms coordinates to track direction of electric field $\mathbf{E}(t)$ of circularly polarized pump pulse}\\
        {\sffamily\footnotesize\bfseries\color{accentteal}\quad $\mathbf{k} \rightarrow \mathbf{k}'$ frame freezing explicitly time-dependent terms}
    };
    \fill[accentteal] ($(step4.north west)+(0.01,-0.01)$) rectangle ($(step4.south west)+(0.12,0.01)$);

    \node[basebox, below=0.4cm of step4] (step5) {
        {\sffamily\bfseries\color{textdark}\small 5. Effective Hamiltonian Approach, Eq.~(\ref{eq:timeindependent})}\\
        {\sffamily\footnotesize\color{textmuted}Constructs an effective static Hamiltonian $H$ and solves the equation in a form of eigenfunctions' decomposition}
    };
    \fill[accentpurple] ($(step5.north west)+(0.01,-0.01)$) rectangle ($(step5.south west)+(0.12,0.01)$);

    \node[basebox, below=0.4cm of step5] (step6) {
        {\sffamily\bfseries\color{textdark}\small 6. Laboratory Frame Back-Transformation, Eq.~(\ref{eq:solution})}\\
        {\sffamily\footnotesize\color{textmuted}Maps solved system back onto native laboratory metrics}\\
        {\sffamily\footnotesize\bfseries\color{accentpink}\quad $\mathbf{k}' \rightarrow \mathbf{k}$ returning variables to real-time dynamics}
    };
    \fill[accentpink] ($(step6.north west)+(0.01,-0.01)$) rectangle ($(step6.south west)+(0.12,0.01)$);

    
    \node[basebox, minimum width=3.9cm, text width=3.4cm, below left=0.7cm and -3.9cm of step6, minimum height=2.0cm] (term1) {
        {\sffamily\bfseries\color{textdark}\footnotesize Interband Coupling}\\
        \vspace{2pt}
        {\sffamily\scriptsize\color{textmuted} Confirms results of the previous studies, Eq.~(\ref{eq:first_term})\\
        \itshape Refs: Slobodeniuk et al., \cite{Slobodeniuk2022, Slobodeniuk2023}}
    };
    \fill[textmuted] ($(term1.north west)+(0.01,-0.01)$) rectangle ($(term1.south west)+(0.10,0.01)$);

    \node[basebox, minimum width=3.9cm, text width=3.4cm, below right=0.7cm and -3.9cm of step6, fill=novelbg, draw=novelgreen!30, minimum height=2.0cm] (term2) {
        {\sffamily\bfseries\color{noveltext}\footnotesize Intraband Dynamics}\\
        \vspace{2pt}
        {\sffamily\scriptsize\color{noveltext} Accounts the contribution of intraband motion, Eq.~(\ref{eq:second_term}) \\
        \bfseries \color{novelgreen}$\star$ NOVEL RESULT}
    };
    \fill[novelgreen] ($(term2.north west)+(0.01,-0.01)$) rectangle ($(term2.south west)+(0.10,0.01)$);

    \draw[arrow] (step1) -- (step2);
    \draw[arrow] (step2) -- (step3);
    \draw[arrow] (step3) -- (step4);
    \draw[arrow] (step4) -- (step5);
    \draw[arrow] (step5) -- (step6);
    
    \path (step6.south) --++ (0,-0.25) coordinate (fork);
    \draw[line width=1.0pt, color=textmuted] (step6.south) -- (fork);
    \draw[arrow] (fork) -| (term1.north);
    \draw[arrow] (fork) -| (term2.north);

\end{tikzpicture}

\section{Hamiltonian formulation} 
\label{sec:two_band_model}

\begin{figure}
\includegraphics[width = 0.4\textwidth]{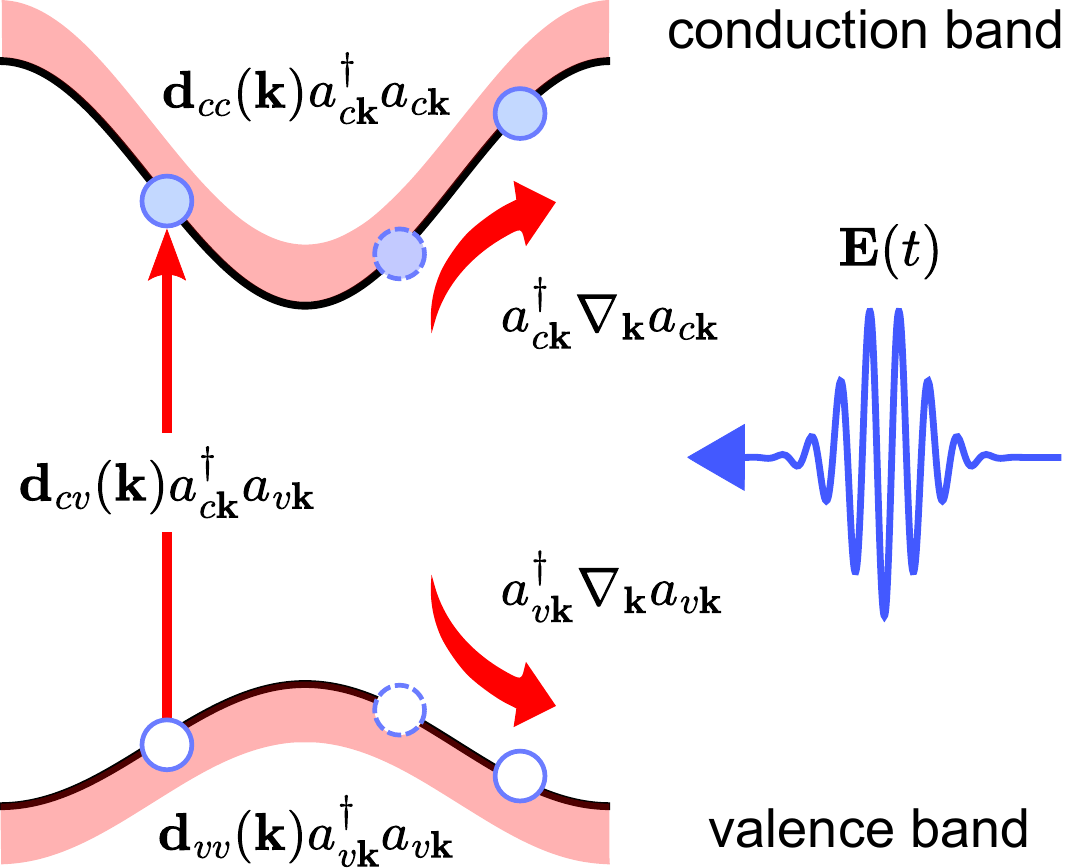}
\caption{\label{fig:fig_1} Valence and conduction bands are electromagnetically coupled by: (i) an interband term, responsible 
for interband transitions (vertical arrow); (ii) intraband motion terms, responsible for electric-field-induced drag of the
charge carriers within the bands (curved arrows); and (iii) intraband Berry connection terms, which effectively modify the 
position of the bands (pink highlighted area).}
\end{figure}

We examine the optical properties of a two-dimensional crystal within the two-band approximation. 
This means that we restrict our consideration only to the processes that appear in the conduction ($c$) and valence ($v$) bands of the semiconductor. 
The dispersion relations of the aforementioned bands as functions of 
two-dimensional quasimomentum vector $\mathbf{k}=(k_x,k_y)$ in the first Brillouin zone (BZ) are $E_{c}(\mathbf{k})$ and $E_{v}(\mathbf{k})$, respectively.
The Hamiltonian of the bands' excitations can be written in the second quantized form as \cite{Slobodeniuk2022,Haug2009,Kira2011} 
\begin{align}
\mathcal{H}_0=\int_\text{BZ} d^2\mathbf{k}\,[E_c(\mathbf{k})a_{c\mathbf{k}}^\dagger a_{c\mathbf{k}}+ 
E_v(\mathbf{k})a_{v\mathbf{k}}^\dagger a_{v\mathbf{k}}].
\end{align}      
Here $a_{c\mathbf{k}}^\dagger (a_{v\mathbf{k}}^\dagger)$ and 
$a_{c\mathbf{k}} (a_{v\mathbf{k}})$ are the creation and annihilation operators of the conduction (valence)
band electrons, which satisfy the standard anticommutation rules 
\begin{align}
\{a_{n\mathbf{k}}, a_{m\mathbf{q}}^\dagger\}=\delta_{nm}\delta(\mathbf{k}-\mathbf{q}), \quad n,m \in c,v.  
\end{align} 
The in-plane time-dependent electric field $\mathbf{E}(t)=(\mathcal{E}_x(t),\mathcal{E}_y(t))$
acts on the electrons' motion in the crystal. The corresponding interaction Hamiltonian reads 
(see the detailed derivation in the appendix \ref{app:interaction}) 
\begin{align}
\label{eq:interaction}
\mathcal{H}_\text{int}=-ie&\mathbf{E}(t)\int_\text{BZ} d^2\mathbf{k}\, [a_{c\mathbf{k}}^\dagger 
\nabla_\mathbf{k}a_{c\mathbf{k}}+a_{v\mathbf{k}}^\dagger \nabla_\mathbf{k}a_{v\mathbf{k}}]
\nonumber \\
-e&\mathbf{E}(t)\int_\text{BZ} d^2\mathbf{k}\,
[\mathbf{d}_{cv}(\mathbf{k})a_{c\mathbf{k}}^\dagger a_{v\mathbf{k}}+
\mathbf{d}_{vc}(\mathbf{k})a_{v\mathbf{k}}^\dagger a_{c\mathbf{k}}]
 \nonumber \\
-e&\mathbf{E}(t)\int_\text{BZ} d^2\mathbf{k}\,
[\mathbf{d}_{cc}(\mathbf{k})a_{c\mathbf{k}}^\dagger a_{c\mathbf{k}}+\mathbf{d}_{vv}(\mathbf{k})a_{v\mathbf{k}}^\dagger a_{v\mathbf{k}}].
\end{align}
Here $e<0|$ is an elementary electron charge, $\nabla_\mathbf{k}=(\partial_{k_x},\partial_{k_y})$ 
is the nabla operator in the momentum space, and $\mathbf{d}_{cv}(\mathbf{k})=\mathbf{d}^*_{vc}(\mathbf{k})$
is the interband transition dipole moment, see details in Refs.~[\onlinecite{Slobodeniuk2022,Slobodeniuk2023,Grzeszczyk2021}], 
while $\mathbf{d}_{vv}(\mathbf{k}), \mathbf{d}_{cc}(\mathbf{k})\in \mathbb{R}$ are Berry connections of the valence and continuity bands, respectively. 
The first term in the Hamiltonian is responsible for the intraband motion of the charge carriers 
in the conduction and valence bands, respectively. This term is non-negligible 
in the case of constant or low-frequency $\omega$ electric field $\mathbf{E}(t)$, i.e. $\hbar\omega \ll E_c(\mathbf{k})-E_v(\mathbf{k})$.       
The second term in the Hamiltonian corresponds to the interband coupling.  
It plays a dominant role when the frequency $\omega$ of the applied electric field  $\mathbf{E}(t)$ satisfies 
near-resonant condition  $\hbar\omega \approx E_c(\mathbf{k})-E_v(\mathbf{k})$. 
The last term describes the influence of the electric field on the carriers in the valence and conduction bands, via the corresponding Berry connections 
$\mathbf{d}_{vv}(\mathbf{k}), \mathbf{d}_{cc}(\mathbf{k})$, which, in particular, are responsible for the anomalous Hall effect in crystalline 
solids\cite{Karplus1954,Kohn1957,Adams1959}. 
In other cases, these terms are usually ignored due to their smallness in weak electric fields \cite{Katsnelson1989, Keldysh1994}. 
The detailed analysis of the Berry connections is presented in appendix \ref{app:berry_connection}. 
A sketch of the band structure with the intra- and interband coupling terms is presented in Fig.~\ref{fig:fig_1}. 

The Coulomb interaction between the carries in the crystal is defined by the Hamiltonian \cite{Haug2009,Kira2011,footnote2}
\begin{align}
\mathcal{H}_\text{C}=&\frac12 \int_\text{BZ}  d^2\mathbf{q}\,V(\mathbf{q})  \times \nonumber \\
\times&  
\sum_{nm=c,v} \iint_\text{BZ} d^2\mathbf{k}\,d^2\mathbf{k}'\, 
a_{n\mathbf{k}'+\mathbf{q}}^\dagger a_{m\mathbf{k}-\mathbf{q}}^\dagger
a_{m\mathbf{k}}a_{n\mathbf{k}'}. 
\end{align} 
Here $V(\mathbf{q})$ is the Fourier transform of the Coulomb potential $V(\mathbf{r})$ 
between two electrons with in-plane distance $r=|\mathbf{r}|$ in a two-dimensional crystal   
\begin{align}
V(\mathbf{q})=\frac{1}{(2\pi)^2}\int d^2\mathbf{r}\,V(\mathbf{r})e^{-i\mathbf{q\cdot r}}.
\end{align}
The shape of the function $V(\mathbf{r})$ depends on many parameters;
see details in Refs.~[\onlinecite{Molas2019,Kipczak2023,Slobodeniuk2023_1,Kapuscinski2024}].  
In our study, we focus on the simplest case of the Rytova-Keldysh potential \cite{Cudazzo2011,Keldysh1979,Rytova1967} 
\begin{align}
V_\text{RK}(\mathbf{r})=\frac{\pi e^2}{2r_0}
\Big[\text{H}_0\Big(\frac{\varepsilon r}{r_0}\Big)-Y_0\Big(\frac{\varepsilon r}{r_0}\Big)\Big],
\end{align}    
which is widely used to describe the Coulomb interaction of few-body complexes S-TMD monolayers.

\section{Semiconductor Bloch equations}
\label{sec:sbe}

We consider the full Hamiltonian $\mathcal{H}=\mathcal{H}_0+\mathcal{H}_\text{int}+\mathcal{H}_\text{C}$
in the Heisenberg picture and  derive the equations of motion for the creation $a_{n\mathbf{k}}^\dagger$ and
annihilation $a_{n\mathbf{k}}$ operators, $n=c,v$. Using these equations and the averaging
procedure $\langle\dots\rangle$ with respect to the initial state of the system, we evaluate the dynamical 
equations for the polarization $P_{vc}(\mathbf{k},t)\equiv\langle a_{c\mathbf{k}}^\dagger(t) a_{v\mathbf{k}}(t)\rangle$, 
and occupation numbers of conduction $N_c(\mathbf{k},t)\equiv\langle a_{c\mathbf{k}}^\dagger(t) a_{c\mathbf{k}}(t)\rangle$ 
and valence $N_v(\mathbf{k},t)\equiv\langle a_{v\mathbf{k}}^\dagger(t) a_{v\mathbf{k}}(t)\rangle$ bands. 
They are called Semiconductor Bloch Equations (SBE), and have in the commonly used Hartree-Fock approximation the following form
\begin{align}
\label{eq:pcv}
\mathcal{D}P_{vc}(\mathbf{k},t)=&-\frac{i}{\hbar}\Big[\mathcal{E}_c(\mathbf{k},t)-\mathcal{E}_v(\mathbf{k},t)\Big]
P_{vc}(\mathbf{k},t)+\nonumber \\&+
\frac{i}{\hbar}\Omega_{cv}(\mathbf{k},t)\Big[N_c(\mathbf{k},t)-N_v(\mathbf{k},t)\Big], 
\\[1ex]
\label{eq:nc}
\mathcal{D}N_c(\mathbf{k},t)=&-\frac{2}{\hbar}\text{Im}[\Omega_{vc}(\mathbf{k},t)P_{vc}(\mathbf{k},t)]
\\[1ex]
\label{eq:nv}
\mathcal{D}N_v(\mathbf{k},t)=&\frac{2}{\hbar}\text{Im}[\Omega_{vc}(\mathbf{k},t)P_{vc}(\mathbf{k},t)]
\end{align}
Here we introduced the material (Lagrangian) derivative, $\mathcal{D}$,  
\begin{align}
\label{eq:lagrangian_derivative}
\mathcal{D}=\frac{\partial}{\partial t}-\frac{|e|}{\hbar}\mathbf{E}(t)\cdot
\nabla_\mathbf{k}, 
\end{align}
renormalized single-particle band energies $\mathcal{E}_n(\mathbf{k},t)$, $n\in c,v$, 
\begin{align}
\mathcal{E}_n(\mathbf{k},t)=E_n(\mathbf{k})+|e|\mathbf{E}(t)\cdot\mathbf{d}_{nn}(\mathbf{k})-\int_\text{BZ}d^2\mathbf{q}\, V(\mathbf{k}-\mathbf{q})N_n(\mathbf{q},t),
\end{align}
and Rabi energy function, $\Omega_{cv}(\mathbf{k},t)=\Omega_{vc}^*(\mathbf{k},t)$,
\begin{align}
\Omega_{cv}(\mathbf{k},t)=|e|\mathbf{E}(t)\cdot\mathbf{d}_{cv}(\mathbf{k})-
\int_\text{BZ} d^2\mathbf{q}\,V(\mathbf{k}-\mathbf{q})P_{vc}(\mathbf{q},t).
\end{align} 

The SBE can be simplified by considering the sum $N_c(\mathbf{k},t)+ N_v(\mathbf{k},t)$ and the difference 
$N_c(\mathbf{k},t)-N_v(\mathbf{k},t)$ of the occupation numbers. 
The corresponding dynamical equations then read
\begin{align}
\mathcal{D}[N_c(\mathbf{k},t)+N_v(\mathbf{k},t)]&=0, \\
\mathcal{D}[N_c(\mathbf{k},t)-N_v(\mathbf{k},t)]&= 
-\frac{4}{\hbar}\text{Im}[\Omega_{vc}(\mathbf{k},t)P_{vc}(\mathbf{k},t)].
\end{align}
The first equation can be solved with the help of the method of characteristics \cite{Courant1962}. The solution reads
\begin{align}
N_c(\mathbf{k},t)+N_v(\mathbf{k},t)=&\sum_{n=c,v}\mathcal{N}_n\Big[\mathbf{k}+\frac{|e|}{\hbar}
\int_0^t dt'\mathbf{E}(t')\Big],
\end{align} 
where $\mathcal{N}_c(\mathbf{k})=N_c(\mathbf{k},0)$ and 
$\mathcal{N}_v(\mathbf{k})=N_v(\mathbf{k},0)$ are the initial occupation numbers of the 
conduction and valence bands with momentum $\mathbf{k}$. This general solution is simplified 
for the special case of the fully occupied valence band $\mathcal{N}_v(\mathbf{k})=1$, and the empty conduction band
$\mathcal{N}_c(\mathbf{k})=0$. In this case  $N_c(\mathbf{k},t)+N_v(\mathbf{k},t)=1$, 
which is nothing but the charge conservation law. Then, introducing the notation $n(\mathbf{k},t)=N_c(\mathbf{k},t)$, we obtain $N_v(\mathbf{k},t)=1-N_c(\mathbf{k},t)=1-n(\mathbf{k},t)$. 
Taking into account that $1-N_v(\mathbf{k},t)$ is an occupation number of holes in the valence band, 
we conclude that $n(\mathbf{k},t)$ is the number of electron-hole pairs with momentum 
$\mathbf{k}$ in the moment of time $t$. Therefore, $n(\mathbf{k},t)\in[0,1]$ \cite{footnote1}. 

The equation for the difference $N_c(\mathbf{k},t)- N_v(\mathbf{k},t)=2n(\mathbf{k},t)-1$ can be rewritten as   
\begin{align}
\label{eq:equation_n}
\mathcal{D}n(\mathbf{k},t)=-
\frac{2}{\hbar}\text{Im}[\Omega(\mathbf{k},t)P(\mathbf{k},t)],
\end{align}
where $P(\mathbf{k},t)\equiv P_{vc}(\mathbf{k},t)$, and 
\begin{align}
\Omega(\mathbf{k},t)=|e|\mathbf{E}(t)\cdot\mathbf{d}_{vc}(\mathbf{k})-
\int_\text{BZ} d^2\mathbf{q}\, V(\mathbf{k}-\mathbf{q})\,P^*(\mathbf{q},t).   
\end{align}
The equation of motion for the polarization $P(\mathbf{k},t)$ reads 
\begin{align}
\label{eq:equation_p}
\mathcal{D}P(\mathbf{k},t)=-\frac{i}{\hbar}e(\mathbf{k},t)
P(\mathbf{k},t)+\frac{i}{\hbar}\Omega^*(\mathbf{k},t)[2n(\mathbf{k},t)-1]. 
\end{align}
Here, we have introduced the energy parameter, $e(\mathbf{k},t)\equiv\mathcal{E}_c(\mathbf{k},t)-\mathcal{E}_v(\mathbf{k},t)$,  
\begin{align}
\label{eq:energy_parameter}
e(\mathbf{k},t)=E_\text{g}(\mathbf{k})-2\int_\text{BZ}d^2\mathbf{q}\, V(\mathbf{k}-\mathbf{q})n(\mathbf{q},t),
\end{align}
with the renormalized band gap (see details in Ref.~[\onlinecite{Slobodeniuk2022})]
\begin{align}
\label{eq:band_gap}
E_\text{g}(\mathbf{k})= &
E_c(\mathbf{k})-E_v(\mathbf{k})+ |e|\mathbf{E}(t)[\mathbf{d}_{cc}(\mathbf{k})-\mathbf{d}_{vv}(\mathbf{k})]+
\nonumber \\ +&
\int d^2\mathbf{q}\, V(\mathbf{q}).
\end{align}
To obtain the last term, we have extended the integration domain from the BZ to the full 2D momentum space and changed the integration variables $\mathbf{k}-\mathbf{q}\rightarrow \mathbf{q}$. 

Using the equations of motion for $n(\mathbf{k},t)$ (\ref{eq:equation_n}) and $P(\mathbf{k},t)$ (\ref{eq:equation_p}) we observe that
\begin{align}
\mathcal{D}\Big[4|P(\mathbf{k},t)|^2+(2n(\mathbf{k},t)-1)^2\Big]=0.
\end{align}
Using the method of characteristics again, we thus obtain 
\begin{align}
4|P(\mathbf{k},t)|^2+(2n(\mathbf{k},t)-1)^2=f\Big[\mathbf{k}+
\frac{|e|}{\hbar}\int_0^t dt'\mathbf{E}(t')\Big], 
\end{align}
where $f(\mathbf{p})$ is a function determined by the initial conditions. 
Since at $t=0$ $n(\mathbf{k},t)=0$ and $P(\mathbf{k},t)=0$, therefore $f(\mathbf{k})=1$, and hence 
\begin{align}
\label{eq:conservation_law}
4|P(\mathbf{k},t)|^2+(2n(\mathbf{k},t)-1)^2=1.
\end{align}  
This integral of motion is a generalization of the similar result previously obtained for the problems, where only inter-band coupling was taken into account, while the intra-band ones were neglected.  
It is advantageous to use this conservation law and the differential equation for the polarization $P(\mathbf{k},t)$ only instead of
the set of coupled differential equations for $P(\mathbf{k},t)$ and $n(\mathbf{k},t)$.  

\section{SBE for S-TMD monolayer}
\label{sec:sbe_2d}

\begin{figure}
\includegraphics[width = 0.4\textwidth]{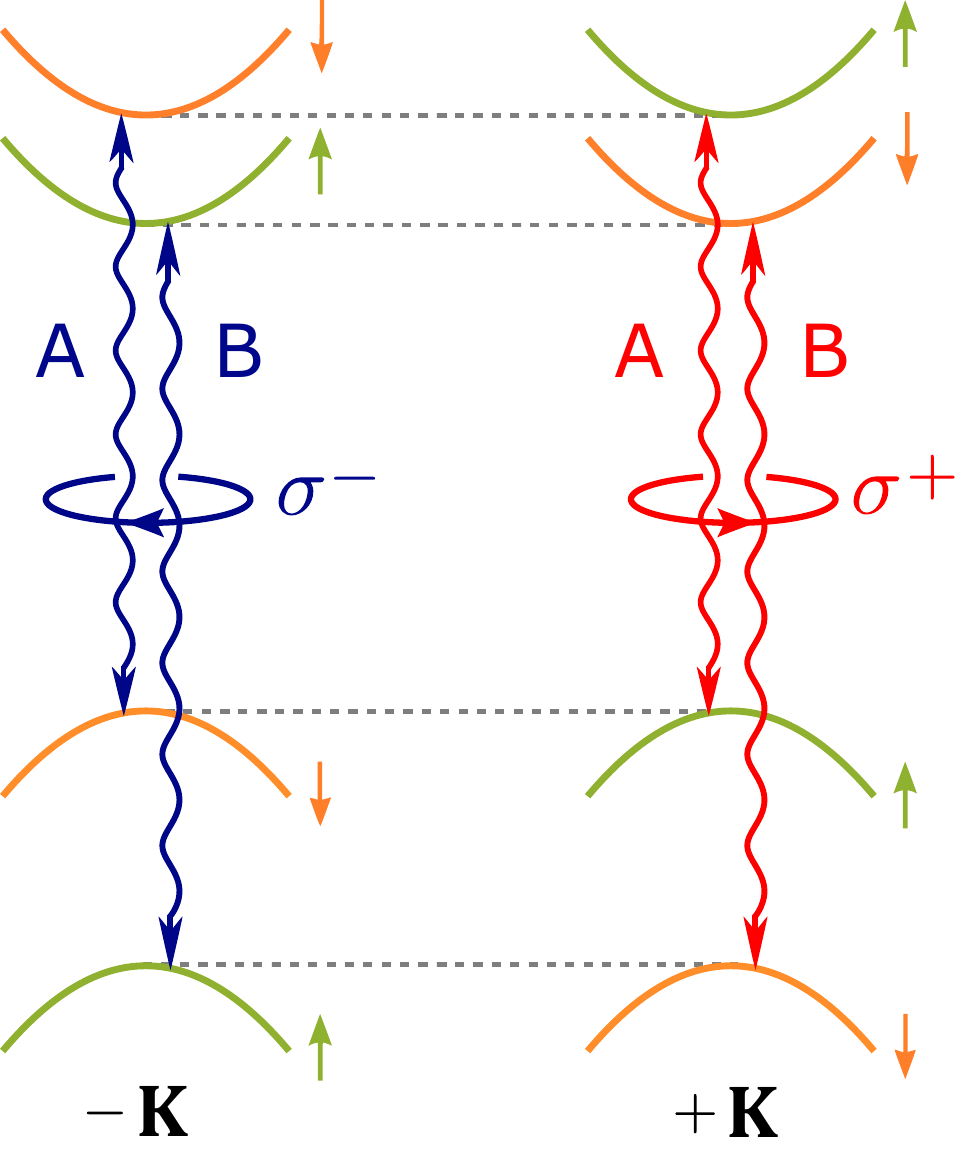}
\caption{\label{fig:fig_2} Band structure of darkish S-TMD (such as WSe$_2$ or WS$_2$) monolayers in  $\pm \mathbf{K}$ valleys. 
Green and orange solid curves show spin-up and spin-down bands in the vicinity of $+\mathbf{K}$ and $-\mathbf{K}$ points. 
Double-headed wavy arrows represent the structure of the interband electromagnetic coupling in each valley. The blue/red color of the arrows indicates 
the $\sigma^-$/$\sigma^+$ circular polarization of the light, which couples exclusively the bands in $-\mathbf{K}/+\mathbf{K}$ point. 
This property is manifested in the structure of the transitions dipole moments $\mathbf{d}_{cv}(-\mathbf{K})/\mathbf{d}_{cv}(+\mathbf{K})$ 
in the corresponding valleys. For the particular case of the photons of resonance energies (which match the distance between valance and conduction bands of the same spin), 
this coupling is responsible for the excitonic transitions in $\pm\mathbf{K}$ points. The capital letters A and B label the low- and high-energy excitonic transitions in each valley.
The dashed grey horizontal lines depict the same positions of the conduction and valence bands of opposite spins in opposite valleys,
which is a consequence of time-reversal symmetry of the crystal.  }
\end{figure}

We adapt the obtained SBE equations to the particular case of S-TMD monolayers. 
As was previously observed \cite{Slobodeniuk2022,Slobodeniuk2023}, the dominant contribution to the optical properties of these 2D crystals
appears in the corners of these hexagonal BZ, the so-called $\pm \mathbf{K}$ points / valleys. The band structure of a S-TMD monolayer at
these points is shown in Fig.~\ref{fig:fig_2}.

Taking into account that $+\mathbf{K}$ and $-\mathbf{K}$ points are non-equivalent, 
we write two separate SBEs in each valley independently. Since the in-plane electric field only couples bands with the same spin, and there is a large energy gap between spin-up and spin-down bands, we assume that the SBEs for spin-up and spin-down bands can be treated independently.
Since these equations have the same structure and differ only by material parameters, such as the renormalized bandgaps $E_\text{g}(\mathbf{k})$, and the matrix elements $\mathbf{d}_{nm}(\mathbf{\mathbf{k}})$, we consider spin-up and spin-down cases simultaneously in each $\pm\mathbf{K}$ point. 

To do this we shift the momenta $\mathbf{k}\rightarrow \pm\mathbf{K}+\mathbf{k}$ in the SBE, obtained in the previous section,  at the $\pm\mathbf{K}$ points of the monolayer, respectively. 
Then the equations take the form  
\begin{align}
\mathcal{D}P(\pm\mathbf{K}+\mathbf{k},t)=&-\frac{i}{\hbar}e(\pm\mathbf{K}+\mathbf{k},t)
P(\pm\mathbf{K}+\mathbf{k},t)+\nonumber \\+&
\frac{i}{\hbar}\Omega^*(\pm\mathbf{K}+\mathbf{k},t)[2n(\pm\mathbf{K}+\mathbf{k},t)-1].
\end{align}
Using Eqs.~(\ref{eq:energy_parameter}) and (\ref{eq:band_gap}), and introducing the valley index  
$\tau=\pm1$ for $\pm\mathbf{K}$ point, we approximate the energy parameter 
$e^\tau(\mathbf{k},t)\equiv e(\tau\mathbf{K}+\mathbf{k},t)$ as
\begin{align}
e^\tau(\mathbf{k},t)\approx& E_\text{g}+\frac{\hbar^2\mathbf{k}^2}{2\mu}+|e|\mathbf{E}(t)\cdot\mathbf{D}^\tau(\mathbf{k})- \nonumber \\-&2\int d^2\mathbf{q}\, 
V(\mathbf{k}-\mathbf{q})n^\tau(\mathbf{q},t).
\end{align}  
To derive this result we took into account that the dispersions of the conduction and valence bands 
have extrema in $\pm\mathbf{K}$ points, with the distance $E_\text{g}$ between these extrema. 
$\mu=m_e|m_v|/(m_e+|m_v|)$ is a reduced exciton mass, where $m_e,|m_v|$ are the  effective masses 
of the conduction and valence bands' electrons in $\pm\mathbf{K}$ points, respectively. 
Note that the energy profile of the bands in $\pm\mathbf{K}$ points is the same 
$E_\text{g}(\mathbf{K}+\mathbf{k})=E_\text{g}(-\mathbf{K}-\mathbf{k})$ 
due to time reversal symmetry of the system. We introduced the notations 
$\mathbf{D}^\tau(\mathbf{k})\equiv\mathbf{d}_{cc}(\tau\mathbf{K}+\mathbf{k})-\mathbf{d}_{vv}(\tau\mathbf{K}+\mathbf{k})$, and 
$n^\tau(\mathbf{k},t)\equiv n(\tau\mathbf{K}+\mathbf{k},t)$.  

The Rabi energy, $\Omega^\tau(\mathbf{k},t)\equiv
\Omega(\tau\mathbf{K}+\mathbf{k},t)$, near $\tau\mathbf{K}$ points reads
\begin{align}
\Omega^\tau(\mathbf{k},t)\approx
|e|\mathbf{E}(t)\cdot\mathbf{d}^\tau_{vc}-
\int d^2\mathbf{q}\, V(\mathbf{k}-\mathbf{q})\,P^{\tau*}(\mathbf{q},t),   
\end{align} 
where we replaced the functions $\mathbf{d}_{vc}(\pm\mathbf{K}+\mathbf{k})$ by their values 
$\mathbf{d}_{vc}(\tau\mathbf{K}+\mathbf{k})\rightarrow 
\mathbf{d}_{vc}(\tau\mathbf{K})\equiv \mathbf{d}^\tau_{vc}$ at 
$\pm\mathbf{K}$ points, and introduced $P^\tau(\mathbf{k},t)\equiv P(\tau\mathbf{K}+\mathbf{k},t)$.
Using the above mentioned quantities one can rewrite Eq.~(\ref{eq:conservation_law}) in the form   
\begin{align}
4|P^\tau(\mathbf{k},t)|^2+(2n^\tau(\mathbf{k},t)-1)^2=1.
\end{align}  
Then the equation for $P^\tau(\mathbf{k},t)$ in $\tau$ valley reads 
\begin{align}
\mathcal{D}P^\tau(\mathbf{k},t)=&
-\frac{i}{\hbar}e^\tau(\mathbf{k},t)
P^\tau(\mathbf{k},t)+
\frac{i}{\hbar}\Omega^{\tau*}(\mathbf{k},t)[2n^\tau(\mathbf{k},t)-1],
\end{align}
Note that the Rabi energy simplifies for the case of the electric field of 
the circularly polarized $\sigma^\pm$ light with normal incidence to the monolayer plane
$\mathbf{E}(t)=\mathcal{E}(t)(\cos(\omega t),\pm\sin(\omega t))$, 
namely 
\begin{align}
\Omega^{\tau*}(\mathbf{k},t)=
-\Big[\tau d_{cv}\mathcal{E}^\tau_\pm(t)+
\int d^2\mathbf{q}\, V(\mathbf{k}-\mathbf{q})P^\tau(\mathbf{q})\Big],
\end{align}
where $\mathcal{E}^\tau_\pm(t)=\mathcal{E}(t)\exp(\mp i\tau\omega t)$ and 
$d_{cv}=i|e|v/\bar{E}_\text{g}$. In the latter, $v$ and $\bar{E}_\text{g}$ 
are the Fermi velocity and single-particle band gap in the S-TMD monolayer (see details
in Refs.~[\onlinecite{Slobodeniuk2022},\onlinecite{Slobodeniuk2023}]).   

For the sake of brevity, we define $P^\tau_\mathbf{k}\equiv P^\tau(\mathbf{k},t)$, 
$n^\tau_\mathbf{k}\equiv n^\tau(\mathbf{k},t)$. It is convenient to rewrite the Rabi energy
and the energy parameters in a form   
\begin{align}
\hbar\omega_{R,\mathbf{k}}^\tau=&\tau d_{cv}\mathcal{E}^\tau_\pm(t)+
\int d^2\mathbf{q}\, V(\mathbf{k}-\mathbf{q})P^\tau_\mathbf{q}, \\
\hbar e_k^\tau=&E_\text{g}+\frac{\hbar^2\mathbf{k}^2}{2\mu}+|e|\mathbf{E}(t)\cdot\mathbf{D}^\tau(\mathbf{k})-
2\int d^2\mathbf{q}\, V(\mathbf{k}-\mathbf{q})n^\tau_\mathbf{q}. 
\end{align}
In this notation, the equations of motion for the polarization and the conservation law read 
\begin{align}
\label{eq:equation_for_polarization}
\mathcal{D}P^\tau_\mathbf{k}=&-ie^\tau_k
P^\tau_\mathbf{k}-i\omega^\tau_{R,\mathbf{k}}(2n^\tau_\mathbf{k}-1),
\end{align}
\begin{align}
\label{eq:integral_of_motion}
4|P^\tau_\mathbf{k}|^2+(2n^\tau_\mathbf{k}-1)^2=1.
\end{align} 
The equations coincide with the previously obtained ones \cite{Slobodeniuk2022,Slobodeniuk2023}. The latter do not contain 
the intra-band terms, and can be obtained from (\ref{eq:equation_for_polarization}) and (\ref{eq:integral_of_motion}) 
by taking a limit case $\mathcal{D}\rightarrow \partial/\partial t$ and $\mathbf{D}^\tau(\mathbf{k})\rightarrow 0$.

\section{Polarization in the S-TMD monolayer}
\label{sec:pump}

We consider the equation for $P^\tau_\mathbf{k}$ in the presence of the $\sigma^\pm$ circularly polarized pump pulse,
$\mathbf{E}_\text{p}(t)=\mathcal{E}_\text{p}(t)(\cos(\omega_\text{p}t),
\pm \sin(\omega_\text{p}t))$, with frequency
$\omega_\text{p}$ and time-dependent amplitude envelope $\mathcal{E}_\text{p}(t)$. Then the detailed equation for the polarization (\ref{eq:equation_for_polarization}) reads 
\begin{align}
\label{eq:p_equation}
i\hbar\frac{\partial P^\tau_\mathbf{k}}{\partial t}=&\Big[E_\text{g}+\frac{\hbar^2\mathbf{k}^2}{2\mu}-e\mathbf{E}_\text{p}(t)\cdot\mathbf{D}^\tau(\mathbf{k})
-ie\mathbf{E}_\text{p}(t)\cdot\nabla_\mathbf{k}\Big]P^\tau_\mathbf{k}- \nonumber \\-&
\int d^2\mathbf{q}\, V(\mathbf{k}-\mathbf{q})\Big[P^\tau_\mathbf{q}-
2n^\tau_\mathbf{k}P^\tau_\mathbf{q}+2n^\tau_\mathbf{q}P^\tau_\mathbf{k}\Big]-
\nonumber \\-&
\tau d_{cv}\mathcal{E}_\text{p}(t)e^{\mp i\tau\omega_\text{p}t}(1-2n_\mathbf{k}^\tau).   
\end{align}
Taking into account the conservation law (\ref{eq:integral_of_motion}), we evaluate the parameter 
$n_\mathbf{k}^\tau\in[0,1/2]$ as a function of $|P_\mathbf{k}^\tau|\in[0,1/2]$
\begin{align}
n^\tau_\mathbf{k}=\frac12-\frac12\sqrt{1-4|P^\tau_\mathbf{k}|^2}\approx |P^\tau_\mathbf{k}|^2
+O(|P^\tau_\mathbf{k}|^4),
\end{align} 
The latter approximation provides $5\%\,(10\%)$ relative deviation for 
$|P^\tau_\mathbf{k}|<0.22\,(0.3)$ from the original formula. 
Hence, it is a reasonably good approximation even for relatively large values of the electric
field $\mathcal{E}_\text{p}(t)$. In the considered regime, the non-linear terms in Eq.~(\ref{eq:p_equation}) 
are supposed to be small and can be neglected as a first approximation.
We consider this approximation to provide the analytical solution of the corresponding equation.  
\begin{align}
\label{eq:linearized_equation}
i\hbar\frac{\partial P^\tau_\mathbf{k}}{\partial t}=&\Big[E_\text{g}+\frac{\hbar^2\mathbf{k}^2}{2\mu}-e\mathbf{E}_\text{p}(t)\cdot\mathbf{D}^\tau(\mathbf{k})
-ie\mathbf{E}_\text{p}(t)\cdot\nabla_\mathbf{k}\Big]P^\tau_\mathbf{k}- \nonumber \\-&
\int d^2\mathbf{q}\, V(\mathbf{k}-\mathbf{q})P^\tau_\mathbf{q}-
\tau d_{cv}\mathcal{E}_\text{p}(t)e^{\mp i\tau\omega_\text{p}t}.
\end{align}
This result is a generalization of the Wannier equation in the momentum space for the case where 
the intraband coupling terms are taken into account. 

\subsection{Reduction of the SBE to the time-independent form}

We simplify the equation~(\ref{eq:linearized_equation})
by reducing it to the time-independent form. 
First, using the substitution 
$P^\tau_\mathbf{k}\equiv e^{\mp i\tau\omega_\text{p}t}p^\tau_\mathbf{k}$,
we remove the time dependence from the source, i.e. the last term on the r.h.s. 
of Eq.~(\ref{eq:linearized_equation})
\begin{align}
\label{eq:small_p}
i\hbar\frac{\partial p^\tau_\mathbf{k}}{\partial t}=&\Big[E_\text{g}+\frac{\hbar^2\mathbf{k}^2}{2\mu}
-e\mathbf{E}_\text{p}(t)\cdot\mathbf{D}^\tau(\mathbf{k})
-ie\mathbf{E}_\text{p}(t)\cdot\nabla_\mathbf{k}\Big]p^\tau_\mathbf{k}\mp \nonumber \\ \mp& \tau\hbar\omega_\text{p}p^\tau_\mathbf{k}-
\int d^2\mathbf{q}\, V(\mathbf{k}-\mathbf{q})p^\tau_\mathbf{q}-
\tau d_{cv}\mathcal{E}_\text{p}(t).
\end{align}
Then we eliminate the time dependence from $\mathbf{E}_\text{p}(t)$ in the intraband terms   
by rewriting the equation in co-moving coordinate system with coordinates 
$\mathbf{k}'=(k_x',k_y')$ 
\begin{align}
k_x'=&k_x\cos\alpha-k_y\sin\alpha, \\
k_y'=&k_x\sin\alpha+k_y\cos\alpha,  
\end{align}
where $\alpha=\pm\omega_\text{p}t$ for $\sigma^\pm$ polarized pump pulses, respectively. 

\begin{figure}
\includegraphics[width = 0.35\textwidth]{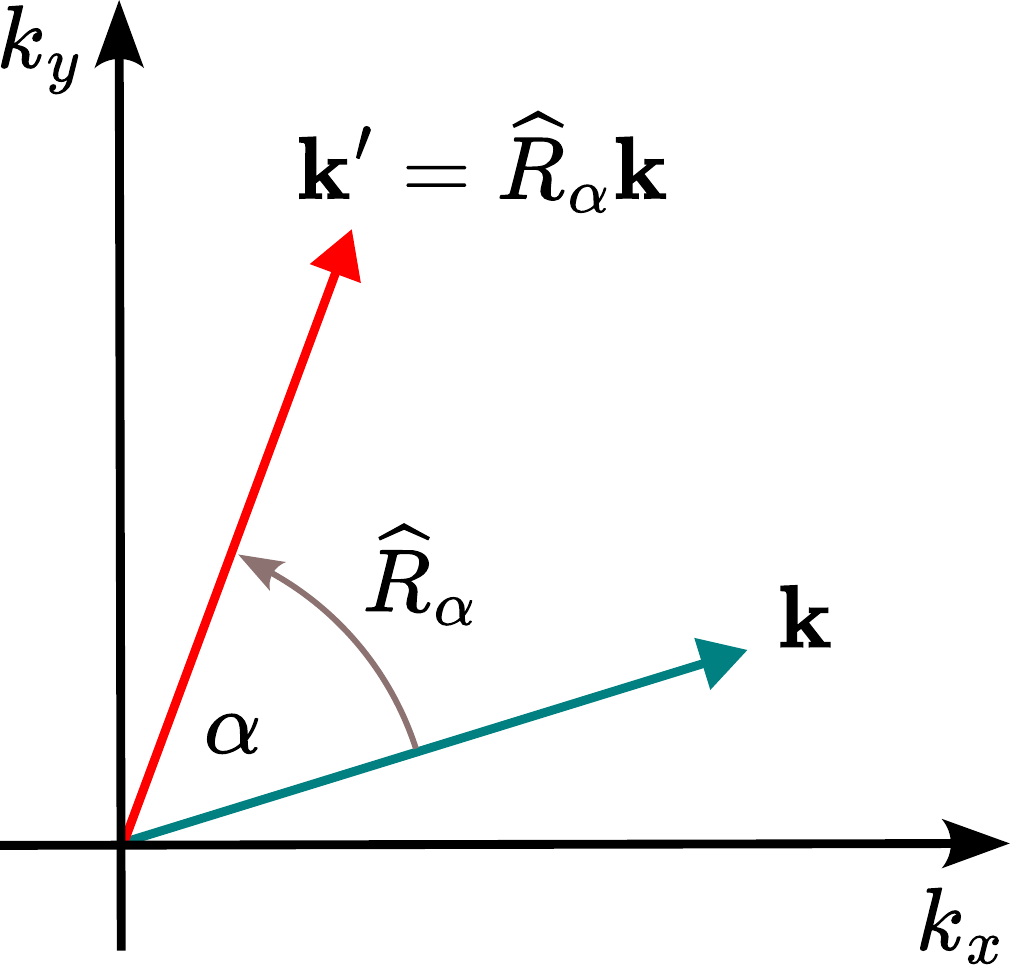}
\caption{\label{fig:fig_3} Sketch of the relation between the two-dimensional vector $\mathbf{k}$ and anticlockwise rotated 
 around the origin of the momentum coordinate system by the angle $\alpha$, 
$\mathbf{k}'=\widehat{R}_\alpha \mathbf{k}$. The symbol $\widehat{R}_\alpha$ represents the rotation operation.}
\end{figure}

The transition to the new coordinate system is realized by a unitary transformation 
$U(\alpha)=\exp(\alpha\frac{\partial}{\partial\theta})$ acting on functions in the momentum space $\mathbf{k}=(k,\theta)$, as
\begin{align}
\label{eq:unitary}
U(\alpha)f(\mathbf{k})=f(k,\theta+\alpha)=
f(\widehat{R}_\alpha\mathbf{k})=f(\mathbf{k}'). 
\end{align}
Here, $\mathbf{k}'=\widehat{R}_\alpha\mathbf{k}$ is a vector $\mathbf{k}$ rotated counterclockwise by the angle $\alpha$ around the origin of the coordinate system.
$\widehat{R}_\alpha$ represents the corresponding rotation operation of the vector demonstrated in Fig.~\ref{fig:fig_3}.  

Note that the action of the operator $U(\alpha)$ on the intraband term simplifies its form
\begin{align}
U(\alpha)\mathbf{E}_\text{p}(t)\cdot\nabla_\mathbf{k}p^\tau_\mathbf{k}=
\mathbf{E}_\text{p}(t)\cdot\nabla_{\mathbf{k}'}p^\tau_{\mathbf{k}'}=
\mathcal{E}_\text{p}(t)\frac{\partial }{\partial k_x}p^\tau_{\widehat{R}_\alpha\mathbf{k}}.
\end{align}   
The latter result prompts us to introduce a new function
$\psi_\mathbf{k}^\tau=U(\alpha)p_\mathbf{k}^\tau=p_{\widehat{R}_\alpha\mathbf{k}}^\tau$.
The function $\psi_\mathbf{k}^\tau$ can be considered as a polarization in the coordinate system 
co-rotating with the electric field of the applied pulse. Moreover, for the particular case of
time-independent magnitude of the pulse $\mathcal{E}_\text{p}(t)=\mathcal{E}_\text{p}$, 
the quasiparticles ``feel'' its static field; in other words we have come to the dynamics of the quasiparticles
in the homogeneous constant field of the ``frozen'' electromagnetic wave.  
The application of the operator $U(\alpha)$ to the term with the Berry connection gives 
\begin{align}
U(\alpha)\mathbf{E}_\text{p}(t)\cdot\mathbf{D}^\tau(\mathbf{k})p^\tau_\mathbf{k}=-\tau \mathcal{E}_\text{p}(t)D(k)k_y \psi^\tau_\mathbf{k},
\end{align}
where we have taken into account the structure of the Berry connection
$\mathbf{D}^\tau(\mathbf{k})=\tau D(k)(-k_y\mathbf{e}_x+k_x\mathbf{e}_y)$, see Appendix~\ref{app:berry_connection}. 

Applying the operator $U(\alpha)$ to Eq.~(\ref{eq:small_p}), and using the definition of the polarization 
$\psi_\mathbf{k}^\tau$ we obtain 
\begin{widetext} 
\begin{align}
\label{eq:full_equation}
i\hbar\frac{\partial \psi^\tau_\mathbf{k}}{\partial t}=\Big[E_\text{g}+\frac{\hbar^2\mathbf{k}^2}{2\mu}+
\tau e\mathcal{E}_\text{p}(t)D(k)k_y 
-ie\mathcal{E}_\text{p}(t)\frac{\partial}{\partial k_x}\pm i\hbar\omega_\text{p}\frac{\partial}{\partial\theta}\mp \tau\hbar\omega_\text{p}\Big]\psi^\tau_\mathbf{k}-
\int d^2\mathbf{q}\, V(\mathbf{k}-\mathbf{q})\psi^\tau_\mathbf{q}-
\tau d_{cv}\mathcal{E}_\text{p}(t).
\end{align} 
\end{widetext}
To modify the integral in this equation, we used
invariance of the potential $V(\mathbf{k})=V(\mathbf{k}')$ and
an area differential $d^2\mathbf{k}=d^2\mathbf{k}'$ under rotation
$\mathbf{k}'=\widehat{R}_\alpha\mathbf{k}$.  
Note that the valley dependence of the equation, the third, sixth, and eighth  of r.h.s. of (\ref{eq:full_equation}), 
originates from $\mathbf{d}_{nm}(\mathbf{k})$-related coupling in the Hamiltonian of the system.  
In contrast, the intraband coupling manifests itself in the fourth and fifth terms of r.h.s. of (\ref{eq:full_equation}).  
In further we consider the case of $\sigma^+$ pulse for brevity. 
The answer for the case of $\sigma^-$ polarization can be found in a similar way.       
We first focus on the case of the time-independent amplitude $\mathcal{E}_\text{p}(t)=\mathcal{E}_\text{p}$. 
It allows us to find the analytical solution of the equation and analyze
the contribution of the intraband terms to this solution.  

\subsection{Construction of the basis states}

In the case of constant amplitude $\mathcal{E}_\text{p}$, the r.h.s. of Eq.~(\ref{eq:full_equation})
doesn't depend on time.
Therefore, it is convenient to use the method of separation of variables to solve this equation. 
Using this approach, we construct the appropriate basis states, which ``catch'' 
the main features of the studied system.  

As a first step, we introduce the set of eigenfunctions $\psi_{nm}(\mathbf{k})$ and eigenvalues $\epsilon_{nm}$ 
which solve the following equation 
\begin{align}
\frac{\hbar^2\mathbf{k}^2}{2\mu}\psi_{nm}(\mathbf{k})-
\int d^2\mathbf{q}\, V(\mathbf{k}-\mathbf{q})\psi_{nm}(\mathbf{q})=\epsilon_{nm}\psi_{nm}(\mathbf{k}).
\end{align}
This equation defines the binding energies and wave functions of the relative motion of the excitons
in a two-dimensional semiconductor \cite{Molas2019,Slobodeniuk2023_1,Kapuscinski2024}. 
The wave functions can be presented in the form     
\begin{align}
\psi_{nm}(\mathbf{k})=\frac{1}{\sqrt{2\pi}}e^{im\theta}\psi_{nm}(k). 
\end{align}
For a discrete part of the spectrum with $\epsilon_{nm}<0$, 
$m=0,\pm1,\pm 2\dots$ is an angular momentum quantum number, and $n=1,2,\dots$ enumerates the states with fixed $m$. 
Using atomic spectroscopy terminology, we name the states with $m=0$, represented by $\psi_{n0}(\mathbf{k})$, as $ns$ states;
$m=\pm1$, represented by $\psi_{n,\pm 1}(\mathbf{k})$, as $np$ states; and $m=\pm2$, represented by $\psi_{n,\pm2}(\mathbf{k})$, as $nd$ states. 
For the case of unbounded (scattering) states with energies $\varepsilon_{nm}>0$, the index $n\equiv(2\hbar^2\varepsilon_{nm}/\mu e^4)^{1/2}$ is a real positive number.  
It can be interpreted as an absolute value of the wave vector of the relative motion of the electron-hole pair.  

Note that $\psi_{nm}(k)=\psi_{n,-m}(k)=\psi_{n|m|}(k)$, because the aforementioned eigenvalue
equation depends only on the absolute value of the angular quantum number $|m|$. 
Hence, the states are doubly degenerate by angular momentum, except the state with $m=0$.  
The eigenfunctions satisfy the relations 
\begin{align}
\int d^2\mathbf{k}\,\psi_{nm}^*(\mathbf{k})\psi_{n'm'}(\mathbf{k})=\delta_{nn'}\delta_{mm'}, \\
\sum_{nm} \psi_{nm}^*(\mathbf{k})\psi_{nm}(\mathbf{k}')=\delta(\mathbf{k}-\mathbf{k}').
\end{align} 
Here $\delta_{nn'}$ is a Kroneker delta and delta function $\delta(n-n')$ for discrete and continuous 
$n,n'$, respectively. 
Analogously, the summation sign means summation over discrete $n$ and integration over continuous ones.  
We use these functions to construct a set of others
\begin{align}
\widetilde{\psi}_{NM}(\mathbf{k})=\sum_{n'm'} C_{NM}^{n'm'}\psi_{n'm'}(\mathbf{k}). 
\end{align}
which solve the full eigenvalue problem, with the eigenvalues $\widetilde{E}_{nm}$, 
\begin{widetext}
\begin{align}
\label{eq:tilde}
\Big[E_\text{g}+\frac{\hbar^2\mathbf{k}^2}{2\mu}+\tau e\mathcal{E}_\text{p}D(k)k_y
-ie\mathcal{E}_\text{p}\frac{\partial}{\partial k_x}+i\hbar\omega_\text{p}\frac{\partial}{\partial\theta}- \tau\hbar\omega_\text{p}\Big]\widetilde{\psi}_{NM}(\mathbf{k}) - 
\int d^2\mathbf{q}\, V(\mathbf{k}-\mathbf{q})\widetilde{\psi}_{NM}(\mathbf{q})=
\widetilde{E}_{nm}\widetilde{\psi}_{NM}(\mathbf{k}).
\end{align}
\end{widetext}
Note that we use the set of parameters $(N,M)$ to enumerate all new solutions. 
These parameters should be understood as the analogs of the principal and angular momentum quantum numbers $(n,m)$.
The solutions $\widetilde{\psi}_{NM}(\mathbf{k})$ are the smooth functions of the parameter $\mathcal{E}_\text{p}$ 
and transform into  $\psi_{NM}(\mathbf{k})$ of the unperturbed problem, in the $\mathcal{E}_\text{p}\rightarrow 0$ limit. 
Hence, $C_{NM}^{nm}\rightarrow \delta_{Nn}\delta_{Mm}$ in the same limit. 
Multiplying both sides of Eq.~(\ref{eq:tilde}) by the function $\psi_{nm}^*(\mathbf{k})$ and integrating it over the momentum space, 
we obtain the secular equation for the coefficients $C_{NM}^{n'm'}$ 
\begin{widetext}
\begin{align}
\label{eq:secular_equation}
\sum_{n'm'}\left[E_{nm}\delta_{nn'}\delta_{mm'}-ie\mathcal{E}_\text{p}
\int d^2\mathbf{k} \psi_{nm}^*(\mathbf{k})
\frac{\partial\psi_{n'm'}(\mathbf{k})}{\partial k_x} +
\tau e\mathcal{E}_\text{p}
\int d^2\mathbf{k} \psi_{nm}^*(\mathbf{k})
D(k)k_y\psi_{n'm'}(\mathbf{k})\right]C_{NM}^{n'm'}=\widetilde{E}_{NM}C_{NM}^{nm}, 
\end{align} 
\end{widetext}
where we introduced $E_{nm}=E_\text{g}+\epsilon_{nm}-\hbar\omega_\text{p}m-\tau\hbar\omega_\text{p}$.
This equation clearly demonstrates that the electric field term mixes the states with different
quantum numbers $(n,m)$ and $(n',m')$ , whereas the terms $\propto \omega_\text{p}$ provide
only the angular momentum shifts to the energies of the states. 

Let us discuss the mixing terms in detail. Taking into account that
\begin{align}
\frac{\partial}{\partial k_x}=\frac12\sum_\pm e^{\pm i\theta}\Big[\frac{\partial}{\partial k}\pm
\frac{i}{k}\frac{\partial}{\partial\theta}\Big], 
\end{align}
and $k_y=k\sin\theta$ we calculate the matrix elements 
\begin{widetext}
\begin{align}
\int d^2\mathbf{k}\,&\psi^*_{nm}(\mathbf{k})
\frac{\partial \psi_{n'm'}(\mathbf{k})}{\partial k_x}=
\frac12\sum_\pm\delta_{m,m'\pm 1} 
\int_0^\infty dk \Big[-k\psi_{n',m\mp1}(k)\frac{d\psi^*_{nm}(k)}{dk}\mp m\psi^*_{nm}(k)\psi_{n',m\mp1}(k)\Big], \\
\int d^2\mathbf{k}\,&\psi^*_{nm}(\mathbf{k})D(k)k_y \psi_{n'm'}(\mathbf{k})=
-\frac{i}{2}\sum_\pm (\pm)\delta_{m,m'\pm 1} 
\int_0^\infty dk k^2D(k) \psi^*_{nm}(k)\psi_{n',m\mp1}(k). 
\end{align}
\end{widetext}
To analyze the system of equations (\ref{eq:secular_equation}) we introduce the infinite vector
$\widetilde{\Psi}_{NM}=[C^{10}_{NM},C^{20}_{NM},\dots C^{11}_{NM},C^{21}_{NM},\dots C^{1,-1}_{NM},C^{2,-1}_{NM},\dots 
C^{1,2}_{NM},C^{2,2}_{NM},\dots]^T$. The components of this vector are ordered according to the following convention. 
First, the coefficients $C_{NM}^{nm}$ corresponding to $m=0$ are listed in 
order of increasing principal quantum number, $n=1,2,3,\dots$. Next, the 
coefficients $C_{NM}^{nm}$ corresponding to $m=1$ are arranged in the same 
manner, followed by the coefficients corresponding to $m=-1$. The same 
procedure is then applied successively for $m=2$, $m=-2$, $m=3$, $m=-3$, 
and so forth.
Using this representation, we rewrite the system of equations (\ref{eq:secular_equation}) in matrix form 
$H\widetilde{\Psi}_{NM}=\widetilde{E}_{NM}\widetilde{\Psi}_{NM}$. 
The Hamiltonian matrix $H$ then reads
\begin{align}
\label{eq:infinite_matrix}
H=\left[
\begin{array}{ccccccc}
H_0 & \Lambda_{01} & \Lambda_{0,-1} & 0 & 0 & 0  & \cdots \\
\Lambda_{01}^\dagger & H_1 &  0 & \Lambda_{12} &  0 & 0  & \cdots   \\
\Lambda_{0,-1}^\dagger & 0 &  H_{-1} & 0 & \Lambda_{-1,-2} &  0 & \cdots  \\ 
0 & \Lambda_{12}^\dagger & 0 & H_2 & 0 & \Lambda_{23} &  \cdots  \\
0 &  0 & \Lambda_{-1,-2}^\dagger & 0 & H_{-2} & 0 & \cdots  \\
0 & 0 & 0 & \Lambda_{23}^\dagger & 0 & H_3 &  \cdots  \\
\vdots & \vdots & \vdots & \vdots & \vdots & \vdots  & \ddots
\end{array}\right].
\end{align}
Here $H_m$ is a diagonal matrix of the energies of the states with angular momentum $m$, which has 
the following matrix elements 
\begin{align}
\label{eq:diagonal_ns}
[H_m]_{nn'}=\delta_{nn'}E_{nm}=\delta_{nn'}[E_\text{g}+\epsilon_{nm}-\hbar\omega_\text{p}m-\tau\hbar\omega_\text{p}].
\end{align} 
Note that we skipped the index $\tau$ from the matrices $H_m$ and $\Lambda_{m,m\pm1}$ for brevity. 
Since the eigenvalue problem has the same structure for both cases $\tau=\pm1$, we solve it in general and then 
restore the index $\tau$ in the final result of the calculation. 

The off-diagonal blocks of the matrix $H$,
$\Lambda_{m,m\pm1}=(ie\mathcal{E}_\text{p}/2)\lambda_{m,m\pm1}$, define the couplings between the states with 
 angular momenta $m$ and $(m\pm1)$, and have the corresponding matrix elements  
\begin{align}
\label{eq:matrix_elements}
[\lambda_{m,m\pm 1}]_{nn'}=&\int_0^\infty\!\!\! dk\, 
\psi_{n',m\pm1}(k)\Big[k\frac{d}{dk}\mp m\Big]\psi^*_{nm}(k) \pm \nonumber \\ \pm& \tau 
\int_0^\infty dk k^2D(k) \psi^*_{nm}(k)\psi_{n',m\pm1}(k).
\end{align}
Note that the distances between the energies of the blocks $H_m$ and $H_{m\pm1}$ are approximately $\hbar\omega_\text{p}$.
We consider the case where this value is large enough, i.e. when the discrete spectra of the states
with different angular momentum are well separated in the energy domain. The case when this assumption is not valid
corresponds to the situations in which the pump pulse induces either interband excitonic transitions or the transitions between inner states of the exciton.
Both cases correspond to near-resonant phenomena that are beyond the scope of the current study.

Having the matrix $H$ we can define the eigenvalue problem $H\widetilde{\Psi}_{NM}=\widetilde{E}_{NM}\widetilde{\Psi}_{NM}$, 
and find the eigenstates $\widetilde{\Psi}_{NM}$ with the corresponding eigenenergies $\widetilde{E}_{NM}$. 
To solve this problem, we developed the perturbative technique by parameter $e\mathcal{E}_\text{p}$. 
We implement it first for the case of the states $\widetilde{\Psi}_{N0}$, 
representing the excitonic states $Ns$ modified by the circularly polarized 
pump pulse. 
For this case, we present the Hamiltonian matrix $H$ in the block form 
\begin{align}
H=\left[
\begin{array}{cc}
H_0 & T \\
T^\dagger & \mathbb{H}   
\end{array}
\right],
\end{align}
where $T=(i e\mathcal{E}_\text{p}/2)[\lambda_{01}, \lambda_{0,-1}, 0, \dots]$, 
and $\mathbb{H}=D+\Lambda$, where $D$ is a diagonal energy matrix 
\begin{align}
D=\left[
\begin{array}{ccccccc}
 H_1 &  0 & 0 &  0 & 0  & \cdots   \\
 0 &  H_{-1} & 0 & 0 & 0  & \cdots  \\ 
 0 & 0 & H_2 & 0 & 0 &  \cdots  \\
 0 & 0 & 0 & H_{-2} & 0 & \cdots  \\
 0 & 0 & 0 & 0 & H_3 &  \cdots  \\
\vdots & \vdots & \vdots & \vdots & \vdots  & \ddots
\end{array}\right],
\end{align}
and off-diagonal matrix of couplings 
\begin{align}
\Lambda=i\frac{e\mathcal{E}_\text{p}}{2}\left[
\begin{array}{cccccc}
 0 &  0 & \lambda_{12} &  0 & 0 & \cdots   \\
 0 &  0 & 0 & \lambda_{-1,-2} & 0  & \cdots  \\ 
 -\lambda^\dagger_{12} & 0 & 0 & 0 & \lambda_{23} & \cdots  \\
 0 & -\lambda^\dagger_{-1,-2} & 0 & 0 & 0 & \cdots  \\
 0 & 0 & -\lambda^\dagger_{23} & 0 & 0 & \cdots  \\
\vdots & \vdots & \vdots & \vdots & \vdots & \ddots
\end{array}\right].
\end{align}  
The matrices $T,\Lambda\propto e\mathcal{E}_\text{p}$, so they can be considered
as electric-field induced perturbation to the main diagonal Hamiltonian matrix. 
Introducing the vector $\widetilde{\Psi}_{N0}=[\Phi_N, \Psi_N]^T$ we write the eigenvalue problem in the form \cite{Novotny2005}
\begin{align}
\label{eq:matrix_secular_equation}
\left[
\begin{array}{cc}
H_0 & T \\
T^\dagger & \mathbb{H}   
\end{array}
\right]
\left[\begin{array}{cc}
\Phi_N \\
\Psi_N
\end{array}\right]=
\widetilde{E}_{N0}\left[\begin{array}{cc}
\Phi_N \\
\Psi_N
\end{array}\right],
\end{align} 
which is equivalent to the pair of equations
\begin{align}
H_0\Phi_N+T\Psi_N=\widetilde{E}_{N0}\Phi_N, \quad T^\dagger\Phi_N+\mathbb{H}\Psi_N=\widetilde{E}_{N0}\Psi_N.  
\end{align} 
Using the second equation we obtain  
\begin{align}
\Psi_N=(\widetilde{E}_{N0}-\mathbb{H})^{-1}T^\dagger\Phi_N.
\end{align}
Substituting this result into the first equation we get 
\begin{align}
\label{eq:}
\Big[H_0+T(\widetilde{E}_{N0}-\mathbb{H})^{-1}T^\dagger\Big]\Phi_N=\widetilde{E}\Phi_N.
\end{align}
This transcendental equation defines the eigenvalue problem, reduced to the subspace of
states with angular momentum $M=0$.  
In this situation the expression in the square brackets can be considered as an effective Hamiltonian, 
whose second term provides a correction to the spectrum of $H_0$.   
The correction term can be expressed in the form of power series in the small parameter 
$\propto e\mathcal{E}_\text{p}/\hbar\omega_\text{p}$. The result in the leading approximation reads
\begin{align}
\label{eq:correction}
T(\widetilde{E}_{N0}-\mathbb{H})^{-1}T^\dagger\approx 
\Big[\frac{e\mathcal{E}_\text{p}}{2}\Big]^2 \sum_{\zeta=\pm1}\lambda_{0\zeta}(\widetilde{E}_{N0}-H_\zeta)^{-1}\lambda^\dagger_{0\zeta}.
\end{align} 
In this case the effective Hamiltonian matrix has a simple form. It contains: i) the diagonal terms $E_{n0}$, corresponding to  
$ns$ excitonic states, see equation (\ref{eq:diagonal_ns}); ii) quadratic in $\mathcal{E}_\text{p}$ diagonal terms which provide the leading 
correction to the spectrum of $ns$ excitonic states; and 3) quadratic in $\mathcal{E}_\text{p}$ off-diagonal 
terms, which provide the higher-order corrections to the spectrum of excitons. 
Since the effective Hamiltonian is evaluated  up to quadratic order in $\mathcal{E}_\text{p}$, 
we need to keep only quadratic terms for the excitons' energies, i.e., we need to keep only the diagonal terms of the effective Hamiltonian. 

\begin{figure}
\includegraphics[width = 0.4\textwidth]{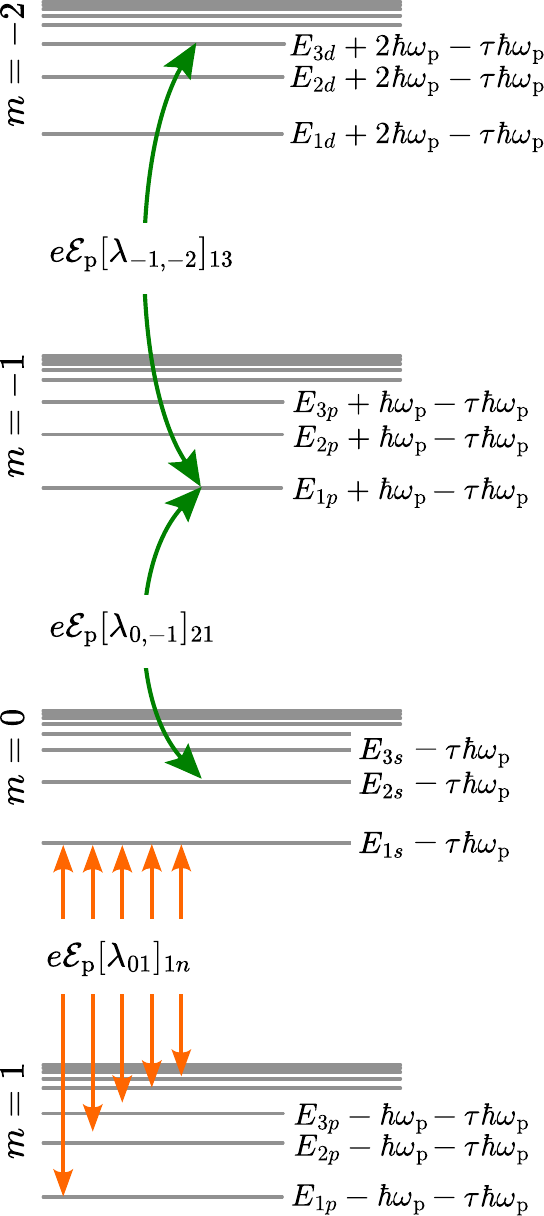}
\caption{\label{fig:fig_4} Hamiltonian structure of the system. Horizontal gray lines represent the relative positions of exciton energies with different angular momenta $m=0,\pm1,\pm2,\dots$. The group of states with closest angular momentum $|m-m'|=1$ are separated approximately by the large energy $\hbar\omega_\text{p}$. The group of states with angular momentum $m$ corresponds 
to the diagonal block $H_m$. The arrows which connect the exciton states with closest angular momenta $|m-m'|=1$ represent the 
electric field $\mathcal{E}_\text{p}$ induced interaction in the monolayer, depicted by matrices $\lambda_{m,m\pm1}$.  }
\end{figure}

Therefore, we can write $\widetilde{E}_{N0}=E_{N0}+\Delta E_{N0}$, where $\Delta E_{N0}$ is a correction to the energy of $Ns$ excitonic states with energy $E_{N0}$, stemming from
(\ref{eq:correction}). Supposing that this correction is small, we replace $\widetilde{E}_{N0}\rightarrow E_{N0}$ in (\ref{eq:correction}) and obtain 
\begin{align}
\label{eq:energies_correction}
\Delta E_{N0}=
\frac{e^2\mathcal{E}^2_\text{p}}{4}\sum_{n',\zeta} 
\frac{|[\lambda_{0\zeta}]_{Nn'}|^2}{E_{Ns}-E_{n'p}+\zeta\hbar\omega_\text{p}}.
\end{align} 
Here we introduced the energies $E_{ns}=E_\text{g}+\epsilon_{n0}$ and $E_{np}=E_\text{g}+\epsilon_{n1}$
of $ns$, and $np$ excitonic states, observed in the experiment.  
The eigenstates, corresponding to the energy $\widetilde{E}_{N0}$, are presented by the vectors $\Phi_N$ with the components $[\Phi_N]_{n'}=\delta_{Nn'}$. 
Using the connection between vectors $\Phi_N$ and $\Psi_N$  we obtain the complementary vector 
\begin{align}
\Psi_N=(E_{N0}-\mathbb{H})^{-1}T^\dagger\Phi_N,
\end{align} 
which together with the previous one form the eigenvector of the full Hamiltonian $[\Phi_N, \Psi_N]^T$, with the energy $\widetilde{E}_{N0}=E_{N0}+\Delta E_{N0}$. 
The calculation gives nonzero values for vector $\Psi_N$ only for $m=\pm1$ states
\begin{align}
[\Psi_N]_{n',\pm 1}=&\Big[\frac{1}{E_{N0}-\mathbb{H}}T^\dagger\Phi_N\Big]_{n',\pm1}=
i\frac{e\mathcal{E}_\text{p}}{2}\frac{[\lambda^\dagger_{0,\pm1}]_{n'N}}{E_{n'p}-E_{Ns}\mp\hbar\omega_\text{p}}, 
\end{align}
where (see the expression for the matrix elements in Eq.~(\ref{eq:matrix_elements})
\begin{align}
[\lambda^\dagger_{0,\pm1}]_{n'N}=[\lambda_{0,\pm1}]_{Nn'}^*=&
\int_0^\infty dk\, k\, \psi_{n'1}^*(k)\frac{d\psi_{N0}(k)}{dk}\pm \nonumber \\ \pm& \tau 
\int_0^\infty dk k^2D(k) \psi^*_{n'1}(k)\psi_{N0}(k).
\end{align}
Therefore, the eigenfunction of the modified $Ns$ states has the form 
\begin{align}
\label{eq:new_functions}
\widetilde{\psi}_{N0}(\mathbf{k})=Z^{-1}_{N0}\Big[\psi_{N0}(\mathbf{k})+i\frac{e\mathcal{E}_\text{p}}{2}
\sum_{n',\pm}\frac{[\lambda^\dagger_{0,\pm1}]_{n'N}\psi_{n',\pm1}(\mathbf{k})}{E_{n'p}-E_{Ns}\mp\hbar\omega_\text{p}} 
\Big], 
\end{align}
with the normalization coefficient 
\begin{align}
\label{eq:norma}
Z_{N0}^2=1+\frac{e^2\mathcal{E}^2_\text{p}}{4}
\sum_{n',\pm}\frac{[\lambda_{0,\pm1}]_{Nn'}[\lambda^\dagger_{0,\pm1}]_{n'N}}{(E_{n'p}-E_{Ns}\mp\hbar\omega_\text{p})^2}. 
\end{align}
The obtained results (\ref{eq:energies_correction}) and (\ref{eq:new_functions}) can be understood with the help of the following graphical scheme of the full infinite Hamiltonian (\ref{eq:infinite_matrix}), depicted in Fig.~\ref{fig:fig_4}. 
Here, the gray lines represent the energy levels of the Hamiltonian (\ref{eq:infinite_matrix}). These levels are grouped into sets corresponding to bound states with different angular momenta, $m=0,\pm1,\pm2,\dots$. Each set corresponds to a diagonal block $H_m$ of the Hamiltonian matrix $H$. Neighboring sets are separated approximately by an energy $\hbar\omega_\text{p}$, as follows from Eq.~(\ref{eq:diagonal_ns}). For example, the states with angular momenta $m=0$ and $m=\pm1$, and principal quantum numbers $n$ and $n'$, respectively, are separated by the energy
\[
\delta E_{nn'} = \left| E_{ns} - E_{n'p} \pm \hbar\omega_\text{p} \right|.
\]
Therefore, these groups of levels remain well separated if $\delta E_{nn'}>|e|[\lambda^\dagger_{0,\pm1}]_{nn'}\mathcal{E}_\text{p}$ for all pairs of indices $(n,n')$.

The arrows in Fig.~\ref{fig:fig_4} indicate the electric-field-induced coupling between states with the closest angular momenta, i.e., $|m-m'|=1$. The coupling strength between states with principal quantum numbers $n$ and $n'$, and angular momenta $m$ and $m'$, is determined by the matrix elements $[\lambda_{mm'}]_{nn'}$, corresponding to the off-diagonal blocks $\Lambda_{mm'}$ of the Hamiltonian matrix $H$. The coupling strength scales linearly with the electric field amplitude, $\Lambda_{mm'}\propto \mathcal{E}_\text{p}$.

The orange arrows in Fig.~\ref{fig:fig_4} represent the coupling between the $1s$ state and various $np$ states. This interaction mixes the $1s$ and $np$ states. Consequently, within the first-order perturbation theory, the admixture of the $np$ states into the $1s$ state is proportional to $\mathcal{E}_\text{p}$. At the same order, the energy of the perturbed $1s$ state acquires a quadratic correction proportional to $\mathcal{E}_\text{p}^2$. These corrections correspond to Eqs.~(\ref{eq:energies_correction}) and (\ref{eq:new_functions}).

The double-headed green arrows illustrate the coupling between states with angular momenta $m=0$ and $m=-2$. This coupling arises as a second-order process mediated by virtual $n'p$ states. For instance, the green arrows in Fig.~\ref{fig:fig_4} correspond to the coupling between the $2s$ and $3d$ states through a virtual $1p$ state. As a result, the admixture of $n''d$ states into the $ns$ states scales as $\mathcal{E}_\text{p}^2$, while the corresponding energy correction scales as $\mathcal{E}_\text{p}^4$. Since this contribution is parametrically small, it is neglected in the present analysis. We therefore conclude that the dominant contribution to the energy shifts and wave-function modification of the $Ns$ excitonic states originates from the coupling to $n'p$ states within the first-order perturbation theory in the electric field $\mathcal{E}_\text{p}$. This conclusion is reflected in the approximation given in Eq.~(\ref{eq:correction}).

The perturbative approach developed here is valid only when groups of states with angular momenta $m$ and $m\pm1$ remain well separated, as illustrated in Fig.~\ref{fig:fig_4}. Under this condition, the expansion parameter 
\begin{align}
e[\lambda^\dagger_{0,\pm1}]_{nn'}\mathcal{E}_\text{p}/(E_{ns}-E_{n'p}\mp\hbar\omega_\text{p})
\end{align}
remains small, allowing one to retain only the leading-order terms in the perturbation series in  
Eqs.~(\ref{eq:correction})-(\ref{eq:norma}). 
Otherwise, when
\begin{align}
|E_{nm}-E_{n',m+1}|\approx \hbar\omega_\text{p},
\end{align}
the photon energy becomes resonant with the excitonic level spacing, leading to optical transitions between the states. Such a resonant regime is beyond the scope of the present study.

An alternative derivation of the results presented in this section is provided in Appendix~\ref{app:derivation}. Finally, the modified basis states with $m\neq0$ can be obtained using the same procedure developed for the $m=0$ states. In particular, the states $\widetilde{\psi}_{N,\pm1}(\mathbf{k})$ contain a small admixture of $\psi_{N0}(\mathbf{k})$. However, this contribution is parametrically small and is therefore neglected in the calculation of the induced polarization.

\subsection{Solution of the SBE equation}

Now we return to the main equation (\ref{eq:full_equation}) with time-independent $\mathcal{E}_{\mathrm{p}}$. To analyze its solutions, we rewrite it in the form
\begin{align}
\label{eq:operator}
\frac{\partial \psi_{\mathbf{k}}^{\tau}}{\partial t}
=\widehat{L}_{\pm}^{\tau}[\psi_{\mathbf{k}}^{\tau}]
-\tau d_{cv}\mathcal{E}_{\mathrm{p}}.
\end{align}
Here we introduced the Hermitian linear integro-differential operator
\begin{widetext}
\begin{align}
\widehat{L}_{\pm}^{\tau}[\psi_{\mathbf{k}}^{\tau}]=\Big[E_{\mathrm{g}}+\frac{\hbar^{2}\mathbf{k}^{2}}{2\mu}+
\tau e\mathcal{E}_{\mathrm{p}}D(k)k_{y}-ie\mathcal{E}_{\mathrm{p}}\frac{\partial}{\partial k_{x}}\pm
i\hbar\omega_{\mathrm{p}}\frac{\partial}{\partial \theta}\mp\tau\hbar\omega_{\mathrm{p}}\Big]
\psi_{\mathbf{k}}^{\tau}-\int d^{2}\mathbf{q}\,V(\mathbf{k}-\mathbf{q})\psi_{\mathbf{q}}^{\tau},
\end{align}    
\end{widetext}
which describes the dynamics of electron excitations in the $\tau$ valley induced by the $\sigma^\pm$ polarized pump pulse.
To construct the general solution of Eq.~(\ref{eq:operator}), we employ the eigenfunctions $\phi_{\pm,\mathbf{k}\beta}^{\tau}$ of the operator $\widehat{L}_{\pm}^{\tau}$, with the corresponding eigenvalues $E_{\pm,\beta}^{\tau}$, labeled by the index $\beta$:
\begin{align}
\widehat{L}_{\pm}^{\tau}[\phi_{\pm,\mathbf{k}\beta}^{\tau}]=E_{\pm,\beta}^{\tau}\phi_{\pm,\mathbf{k}\beta}^{\tau}.
\end{align}
For subsequent analysis, we assume the eigenfunctions to be normalized according to
\begin{align}
\int d^{2}\mathbf{k}\,
\phi_{\pm,\mathbf{k}\beta}^{\tau *}
\phi_{\pm,\mathbf{k}\gamma}^{\tau}
=
\delta_{\beta\gamma}.
\end{align}
Using the Fourier method for solving differential equations, we express the general solution as
\begin{align}
\label{eq:pidstanovka}
\psi_{\pm,\mathbf{k}}^{\tau}(t)=\sum_{\beta}C_{\pm,\beta}^{\tau}(t)e^{-iE_{\pm,\beta}^{\tau}t/\hbar}\phi_{\pm,\mathbf{k}\beta}^{\tau}.
\end{align}
Here we added the subscript ``$\pm$'' in $\psi_{\pm,\mathbf{k}}^{\tau}(t)$ to separate the cases of the $\sigma^{\pm}$ polarization of the pump pulse.
Substituting this Ansatz into Eq.~(\ref{eq:operator}), we obtain
\begin{align}
\label{eq:coefficient}
i\hbar\sum_{\beta}\frac{dC_{\pm,\beta}^{\tau}(t)}{dt}e^{-iE_{\pm,\beta}^{\tau}t/\hbar}
\phi_{\pm,\mathbf{k}\beta}^{\tau}=-\tau d_{cv}\mathcal{E}_{\mathrm{p}}.
\end{align}
The coefficients $C_{\pm,\beta}^{\tau}(t)$ are defined up to an arbitrary constant. Specifically, if $C_{\pm,\beta}^{\tau}(t)$ is a solution of Eq.~(\ref{eq:coefficient}), then $C_{\pm,\beta}^{\tau}(t)+C_{\beta}$, with arbitrary constants $C_{\beta}$, is also a solution. These constants represent the solution of the homogeneous equation
\begin{align}
\frac{\partial \psi_{\mathbf{k}}^{\tau}}{\partial t}=\widehat{L}_{\pm}^{\tau}[\psi_{\mathbf{k}}^{\tau}].
\end{align}
These solutions are independent of the coupling parameter $d_{cv}$. Consequently, they describe states that exist even in the absence of light-matter coupling, i.e., for $d_{cv}=0$.
In the present work, however, we are interested only in the particular solution induced by the coupling term. To determine this solution, we employ the standard adiabatic-switching procedure. Specifically, we replace
\begin{align}
d_{cv}\rightarrow d_{cv}e^{\eta t},
\end{align}
with infinitesimal $\eta>0$, solve Eq.~(\ref{eq:coefficient}) subject to the boundary condition
\begin{align}
C_{\pm,\beta}^{\tau}(-\infty)=0,
\end{align}
and subsequently take the limit $\eta\rightarrow 0$. The resulting solution reads as follows.
\begin{align}
\label{eq:solution}
\psi_{\pm,\mathbf{k}}^{\tau}(t)=\tau d_{cv}\mathcal{E}_{\mathrm{p}}
\sum_{\beta}\phi_{\pm,\mathbf{k}\beta}^{\tau}\frac{
\int d^{2}\mathbf{k}'\,\phi_{\pm,\mathbf{k}'\beta}^{\tau *}}{E_{\pm,\beta}^{\tau}-i0_+}.
\end{align}
The solution (\ref{eq:solution}) exhibits a divergence when $E_{\pm,n}^{\tau}=0$. Physically, this condition corresponds to resonance, where the pump-photon energy $\hbar\omega_{\mathrm{p}}$ matches the energy difference between excitonic states, leading to optically induced transitions.For the parameters considered here, such resonant conditions are not realized, and the obtained solution is regular.

Finally, we note that the solution (\ref{eq:solution}) is time independent
\begin{align}
\frac{\partial \psi_{\pm,\mathbf{k}}^{\tau}}{\partial t}=0,
\end{align}
and therefore satisfies
\begin{align}
\label{eq:redukovane_rivniannia}
\widehat{L}_{\pm}^{\tau}[\psi_{\pm,\mathbf{k}}^{\tau}]=\tau d_{cv}\mathcal{E}_{\mathrm{p}}.
\end{align}
Therefore, the particular solution of the problem corresponds to the Anzats (\ref{eq:pidstanovka}) with time-independent coefficients $C_{\pm,\beta}^{\tau}(t)=C_{\pm,\beta}^{\tau}$.

Using the technique developed above, we provide the detailed solution for the polarization in $\tau$ valley induced by $\sigma^+$- polarized pump pulse for the clarity. The case of $\sigma^-$-polarized light can be obtained analogously. Then using the previously obtained functions $\widetilde{\psi}_{NM}(\mathbf{k})$ as a complete set of orthonormalized eigenfunctions, we present the solution as 
\begin{align}
\psi_\mathbf{k}^\tau=\sum_{NM} C^\tau_{NM}\widetilde{\psi}_{NM}(\mathbf{k}),
\end{align}
with time-independent $C^\tau_{NM}$ coefficients. 
Substituting this Anzats into (\ref{eq:redukovane_rivniannia}), multiplying the result by $\widetilde{\psi}^*_{NM}(\mathbf{k})$ and integrating it over the momentum space we obtain 
\begin{align}
\label{eq:system}
C^\tau_{NM}[E_\text{g}+\epsilon_{NM}-\tau\hbar\omega_\text{p}-&\hbar\omega_\text{p}M+\Delta E_{NM}]=\nonumber \\  
=&\tau d_{cv}\mathcal{E}_\text{p}\int_0^\infty d^2\mathbf{k}\, \widetilde{\psi}^*_{NM}(\mathbf{k}).
\end{align}
The integral is non-zero only for the $M=0$ and $M=\pm 1$ states. This is because only the states $\widetilde{\psi}^*_{N0}(\mathbf{k})$ and $\widetilde{\psi}^*_{N,\pm1}(\mathbf{k})$, contain $Ns$ states $\psi_{N0}(\mathbf{k})$, which give nonzero contribution after integration over $\mathbf{k}$-space. One can show that
\begin{align}
\int_0^\infty &d^2\mathbf{k}\, \widetilde{\psi}^*_{N0}(\mathbf{k}) \propto 1, \\
\int_0^\infty &d^2\mathbf{k}\, \widetilde{\psi}^*_{N, \pm1}(\mathbf{k}) \propto \mathcal{E}_\text{p},
\end{align}
hence, the contribution from the $M=\pm 1$ states to the right hand side of Eq. (\ref{eq:system}) is parametrically small. 
Keeping only the dominant contributions from $M=0$ states, we obtain 
\begin{align}
C^\tau_{NM}=
\delta_{M0}\frac{\tau d_{cv}\mathcal{E}_\text{p}}
{E_{Ns}-\tau\hbar\omega_\text{p}+\Delta E_{N0}}\int_0^\infty d^2\mathbf{k}\, 
\widetilde{\psi}^*_{N0}(\mathbf{k}),
\end{align} 
and hence
\begin{align}
\psi^\tau_\mathbf{k}=\sum_N 
\frac{\tau d_{cv}\mathcal{E}_\text{p}\widetilde{\psi}_{N0}(\mathbf{k})}{E_{Ns}-\tau\hbar\omega_\text{p}+\Delta E_{N0}}
\int_0^\infty d^2\mathbf{k}' \widetilde{\psi}^*_{N0}(\mathbf{k}').
\end{align} 
Here, the summation over $N$ also includes the integration over the continuous parameter.
Substituting the obtained expression for $\widetilde{\psi}^*_{n0}(\mathbf{k}')$ and 
performing integration we obtain 
\begin{align}
\label{eq:timeindependent}
\psi^\tau_\mathbf{k}=&\tau d_{cv}\mathcal{E}_\text{p}
\sum_{N}\frac{\int_0^\infty  dk'k'\,\psi^*_{N0}(k')}
{Z_{N0}^2(E_{Ns}-\tau\hbar\omega_\text{p}+\Delta E_{N0})}\times \nonumber  \\ \times&
\Big[\psi_{N0}(k)+i\frac{e\mathcal{E}_\text{p}}{2}
\sum_{N',\pm} \frac{[\lambda^\dagger_{0,\pm1}]_{N'N}\psi_{N'1}(k)}
{E_{N'p}-E_{Ns}\mp\hbar\omega_\text{p}}e^{\pm i\theta}\Big].
\end{align}
For convenience, we rewrite this result in terms of the coordinate-dependent excitonic wave functions 
\begin{align}
\psi_{nm}(\mathbf{r})\equiv\frac{e^{im\varphi}}{\sqrt{2\pi}}\psi_{n|m|}(r)=\frac{1}{2\pi}\int d^2\mathbf{k}e^{i\mathbf{kr}}\psi_{nm}(\mathbf{k}).
\end{align}
Namely, for the case of $m=0,\pm1$ we obtain 
\begin{align}
\psi_{n0}(k)=&\int_0^\infty dr\,r\,\psi_{n0}(r)J_0(kr), \\
\psi_{n1}(k)=&-i\int_0^\infty dr\,r\,\psi_{n1}(r)J_1(kr).  
\end{align}
Substituting these expressions into the main formula, we obtain 
\begin{align}
\psi^\tau_\mathbf{k}=A^\tau(k)-iB^\tau_+(k)e^{i\theta}-iB^\tau_-(k)e^{-i\theta}.
\end{align}
Here we introduce the parameters $A^\tau(k)$, $B^\tau_+(k)$, and $B^\tau_-(k)$ presented in the following 
\begin{widetext}
\begin{align}
A^\tau(k)=&\tau d_{cv}\mathcal{E}_\text{p}\sum_{N}\frac{\psi^*_{N0}(r=0)}
{Z_{N0}^2(E_{Ns}-\tau\hbar\omega_\text{p}+\Delta E_{N0})}\int_0^\infty dr' r' \psi_{N0}(r')J_0(kr'), \\
B^\tau_\pm(k)=&\tau d_{cv}\frac{e\mathcal{E}_\text{p}^2}{2}
\sum_{N}\frac{\psi^*_{N0}(r=0)}
{Z_{N0}^2(E_{Ns}-\tau\hbar\omega_\text{p}+\Delta E_{N0})}
\sum_{N'}
\frac{\int_0^\infty \int_0^\infty drdr' rr' \psi_{N0}(r)\psi^*_{N'1}(r')\Big[\delta(r-r')\mp\tau D(r,r')\Big]}{E_{N'p}-E_{Ns}\mp\hbar\omega_\text{p}}
\int_0^\infty dr' r' \psi_{N'1}(r')J_1(kr').
\end{align}
\end{widetext}
Here we introduced 
\begin{align}
D(r',r)=\int_0^\infty dk\,k^2D(k)J_1(kr')J_0(kr). 
\end{align}
Using the result for $\psi_\mathbf{k}^\tau$ we finally can obtain the polarization $P_\mathbf{k}^\tau$ induced by $\sigma^+$ polarized electromagnetic pulse in $\tau$ valley of 2D semiconductor. 
Namely, using the definition in Sec.~\ref{sec:pump}~A $P_\mathbf{k}^\tau=e^{-i\tau\omega_\text{p}t}U(-\omega_\text{p}t)\psi^\tau_\mathbf{k}$ together with the definition of the unitary operator $U(\alpha)$, see (\ref{eq:unitary}), we obtain  
\begin{align}
\label{eq:solution}
P_\mathbf{k}^\tau=e^{-i\tau\omega_\text{p}t}
\Big[A^\tau(k)-i\sum _{\pm }B^\tau_\pm(k)e^{\pm i(\theta-\omega_\text{p}t)}\Big].
\end{align}
Note that the time-dependent polarization comprises three distinct components:  a stationary (DC) term, and two oscillating terms evolving at the 
fundamental ($\omega_{\text{p}}$) and second-harmonic ($2\omega_{\text{p}}$) frequencies.  

\subsection{Discussion}
\label{sec:discussion}

The obtained polarization consists of two terms. The first term, 
\begin{align}
\label{eq:first_term}
A^\tau(k)\propto \frac{d_{cv}\mathcal{E}_\text{p}}{E_{Ns}+\Delta E_{Ns}-\tau\hbar\omega_\text{p}},
\end{align}
summarizes the result in the spirit of {\it Elliott formula} \cite{Elliott1957}. 
This contribution originates directly from the inter-band coupling term in the SBE. This result is linear in the applied field $\mathcal{E}_\text{p}$, while the other term is quadratic in the electric field. Therefore, in the limit of weak pump fields, this contribution dominates and reproduces the optical properties previously obtained for the semiconductors, which were mentioned in the Introduction. From this expression we observe that the magnitude of the induced polarization is suppressed by a large exciton energy in the off-resonant regime. Note that the result is different in different valleys, appearing from the specific coupling of light to the bands at opposite $+\mathbf{K}$ and $-\mathbf{K}$ points of the Brillouin zone of the monolayer. The pole structure of the result when $\hbar\omega_\text{p}\approx E_{Ns}$ restores the Elliott formula and manifests the possibility of optical transitions
for the resonant case.

The second term, 
\begin{align}
\label{eq:second_term}
B_\pm^\tau(k)\propto \frac{d_{cv}\mathcal{E}_\text{p}}{E_{Ns}+\Delta E_{Ns}-\tau\hbar\omega_\text{p}}
\frac{e\mathcal{E}_\text{p}[\lambda_{01}]_{NN'}}{E_{N'p} - E_{Ns}\mp \hbar\omega_\text{p}},
\end{align}
can be considered as a second-order contribution by the electric field $B^\tau(k)\propto \mathcal{E}^2_\text{p}$ to the induced polarization in the system. This contribution can be understood as a result of a two-step process in the system. The first step is responsible for the generation of the polarization related to virtual excitonic states $Ns$ with different quantum numbers $N=1,2,3,\dots$. 
The effectiveness of such a process that originates from inter-band coupling is proportional to $d_{cv}\mathcal{E}_\text{p}/(E_{Ns}+\Delta E_{Ns}-\tau\hbar\omega_\text{p})$. The second step appears from the intraband dynamics of the carriers in the $\tau=\pm1$ valleys. Mixes the $Ns$ and $N'p$ excitonic states in the system, where the effectiveness of such mixing is defined by the coupling parameter $[\lambda_{01}]_{NN'}$. 
This process consists of two contributions: i) {\it valley independent}, which originates from intraband motion of the carrriers within the valleys; and ii) {\it valley dependent}, which appears due non-zero values of Berry curvatures of the valence and conduction bands at the corners of the first BZ of the crystal. 

Note that the second term has an additional pole structure, when $\hbar\omega_\text{p}\approx |E_{N'p} - E_{Ns}|$. This situation corresponds to the case of optical transitions between $Ns$ and $N'p$ exciton states for very small photon energies of the pump pulse. In this case, the second contribution to polarization provides the leading correction to the generalized Elliott formula for the domain of small frequencies $\omega_\text{p}$ of the pump pulse. 

Finally, note that our result is valid only for values of the electric field strength $\mathcal{E}_\text{p}$ and frequency $\omega_\text{p}$ of the pump pulse that do not destroy the excitonic states in the material. In this case, we can use the excitonic states $\widetilde{\psi}_{NM}(\mathbf{k})$, modified by the electric field of the pump pulse, in our analysis. Otherwise, the excitonic states, as bound states of electron-hole pairs, cease to exist because of tunneling into the continuum, and the analysis proposed in the current paper is not applicable. This situation is beyond the scope of the present paper and should be considered separately.     

\section{Conclusions}
\label{sec:conclusions}

We have developed a theoretical formalism for the optical response of two-dimensional (2D) semiconductors strongly driven by an intense laser field in the non-resonant regime. This theory, for the first time, properly accounts for the contribution of the {\it intraband coupling} term to the induced polarization in the system. The induced polarization is analyzed within the framework of the Semiconductor Bloch equation, which are reduced to a {\it generalized Wannier equation}. In this formulation, the intraband terms from the SBEs lead to an expression that governs the electric-field-induced coupling between the different states of the electron-hole pairs, which is nothing but the excitonic {\it dynamical Stark effect}. We solved the generalized Wannier equation and obtained an analytical expression for the induced polarization, presenting the result as a power series expansion in a small parameter, proportional to the amplitude of the electric field $\mathcal{E}_\text{P}$ , that naturally appears in the problem.

We found the intraband terms produce an important correction to the polarization, which becomes essential in the off-resonant regime of the applied optical pulse. The significance of this term is especially pronounced in the material's optical response when the interband contribution is equal to zero due to symmetry arguments and/or geometry of the experiment. Therefore, the optical properties of the materials can be modified in the presence of the strong optical pulse even in the cases then the interband light-matter coupling is forbidden. Hence, the intraband contribution is crucial and can't be ignored in the analysis of the modern studies involving highly intensive optical pulses. 

The theoretical methods elaborated in this study provide a complete background for analysis of the optical properties of two-dimensional semiconductors in the presence of non-resonance strong optical fields. We use these methods to develop the theory of the optically-induced excitonic shifts in S-TMD monolayers in the second part of this study.

\begin{acknowledgments}
A.~O.~Slobodeniuk acknowledges the support of the Czech Science Foundation (project GA\v{C}R 26-21965S). 
and thanks L.~Dyachenko, A.~Makukha, S.~ T\'{a}zlar\r{u}, and M.~Koz\'{a}k  for fruitful discussions. 
\end{acknowledgments}

\section*{Author declarations}

{\bf Conflict of Interest} 

The authors have no conflicts to disclose. 

{\bf Author Contributions} 

{\bf A.~O.~Slobodeniuk}: Conceptualization (lead); Formal analysis (equal); Investigation (lead); Methodology (equal); Software (equal); Writing – original draft (lead); Writing – review \& editing (equal). 
{\bf T.~Novotn\'{y}}:  Formal analysis (equal);  Methodology (equal); Software (equal); Writing – review \& editing (equal). 

\section*{Data availability} 

The data that support the findings of this study are available from the corresponding author upon reasonable request.

\appendix

\section{Derivation of the electron-photon interaction Hamiltonian}
\label{app:interaction}

The single-particle Hamiltonian of an electron in a crystal interacting with with a time-dependent electric field of the 
electromagnetic wave reads
\begin{align}
H_\text{int}=e\Phi(\mathbf{r},t)=-e\mathbf{E}(t)\mathbf{r}.
\end{align}
Here, $e=-|e|$ is the electron charge, $\Phi(\mathbf{r},t)$ is the electric potential corresponding to the electric field of the pulse $\mathbf{E}(t)$,  
and $\mathbf{r}$ is a radius vector of the electron in the crystal. We suppose that the electric field is homogeneous throughout the crystal.
In further detail, we consider the case of a two-dimensional crystal, that is $\mathbf{r}=(x,y)$, and the normal incidence of the electromagnetic pulse, 
that is $\mathbf{E}(t)=(\mathcal{E}_x(t),\mathcal{E}_y(t))$. 

In order to write the corresponding interaction Hamiltonian in the second quantized form we introduce the field operator 
\begin{align}
\Psi(\mathbf{r})=\sum_{n} \int_\text{BZ} d^2\mathbf{k} \psi_{n\mathbf{k}}(\mathbf{r})a_{n\mathbf{k}}.
\end{align}
Here $\psi_{n\mathbf{k}}(\mathbf{r})$ is the Bloch function of the $n$th band with the wave-vector $\mathbf{k}$ belonging to the first Brillouin zone (BZ).
It can be represented in the form 
\begin{align}
\psi_{n\mathbf{k}}(\mathbf{r})=\frac{e^{i\mathbf{kr}}}{2\pi}u_{n\mathbf{k}}(\mathbf{r}),
\end{align}
where $u_{n\mathbf{k}}(\mathbf{r})$ is the function invariant under the translations by lattice vectors $\mathbf{a}_1, \mathbf{a}_2$ of the crystal, 
$u_{n\mathbf{k}}(\mathbf{r})=u_{n\mathbf{k}}(\mathbf{r}+\mathbf{a}_1)=u_{n\mathbf{k}}(\mathbf{r}+\mathbf{a}_2)$. 
The Bloch functions satisfy the orthogonality conditions 
\begin{align}
\int_V d^2\mathbf{r}\,\psi^*_{n\mathbf{k}}(\mathbf{r})\psi_{m\mathbf{q}}(\mathbf{r})=\delta_{nm}\delta(\mathbf{k}-\mathbf{q}), 
\end{align}
where integration is performed over the volume of the crystal $V$ which supposed to go to infinity. 
The functions $u_{n\mathbf{k}}(\mathbf{r})$ are orthonormalized within the unit cell of the volume $\Omega$ 
\begin{align}
\frac{1}{\Omega}\int_{\Omega} d^2\mathbf{r}\,u^*_{n\mathbf{k}}(\mathbf{r})u_{m\mathbf{k}}(\mathbf{r})=\delta_{nm}.
\end{align}
$a_{n\mathbf{k}}$ is an annihilation operator of electron's state in the band $n$ with a wave-vector $\mathbf{k}$. The operators $a_{n\mathbf{k}}$ and their Hermitian conjugated ones 
$a^\dagger_{n\mathbf{k}}$ (creation operators) satisfy the following anticommutation relations 
\begin{align}
\{a_{n\mathbf{k}},a^\dagger_{m\mathbf{q}}\}=\delta_{nm}\delta(\mathbf{k}-\mathbf{q}), \quad 
\{a_{n\mathbf{k}},a_{m\mathbf{q}}\}=\{a^\dagger_{n\mathbf{k}},a^\dagger_{m\mathbf{q}}\}=0. 
\end{align}
Then the interaction Hamiltonian in the second quantized form reads 
\begin{align}
\label{eq:h_int}
\mathcal{H}_\text{int}=&|e|\int_V d^2\mathbf{r} \Psi^\dagger(\mathbf{r})\mathbf{E}(t)\mathbf{r}
 \Psi(\mathbf{r})=\nonumber \\=&
 |e|\mathbf{E}(t)\sum_{nm}\iint_\text{BZ} d^2\mathbf{k}d^2\mathbf{q} \,\mathbf{J}_{nm}(\mathbf{k}, \mathbf{q})
a^\dagger_{n\mathbf{k}}a_{m\mathbf{q}},
\end{align}
where 
\begin{align}
\mathbf{J}_{nm}(\mathbf{k}, \mathbf{q})=\int_V d^2\mathbf{r} \psi^*_{n\mathbf{k}}(\mathbf{r})\mathbf{r}\psi_{m\mathbf{q}}(\mathbf{r})
\end{align}
The spatial integral can be evaluated in the following way  
\begin{align}
\label{eq:x_matrix_element}
\mathbf{J}_{nm}(\mathbf{k}, \mathbf{q})=&\frac{1}{(2\pi)^2}
\int_V d^2\mathbf{r}\, e^{-i\mathbf{kr}}u^*_{n\mathbf{k}}(\mathbf{r})\mathbf{r}
e^{i\mathbf{qr}}u_{m\mathbf{q}}(\mathbf{r})=\nonumber \\ =&
\frac{1}{(2\pi)^2}\int_V d^2\mathbf{r}\, e^{-i\mathbf{kr}}u^*_{n\mathbf{k}}(\mathbf{r})\big[-i\nabla_\mathbf{q}
e^{i\mathbf{qr}}\big]u_{m\mathbf{q}}(\mathbf{r})=\nonumber \\=&
-\frac{i}{(2\pi)^2}\int_V d^2\mathbf{r}\, e^{-i\mathbf{kr}}u^*_{n\mathbf{k}}(\mathbf{r})\nabla_\mathbf{q}
\big[e^{i\mathbf{qr}}u_{m\mathbf{q}}(\mathbf{r})\big]+\nonumber \\&+
\frac{i}{(2\pi)^2}\int_V d^2\mathbf{r}\, e^{-i\mathbf{(k-q)r}}u^*_{n\mathbf{k}}(\mathbf{r})\nabla_\mathbf{q}
u_{m\mathbf{q}}(\mathbf{r})=\nonumber \\=&
-i\nabla_\mathbf{q}\int_V d^2\mathbf{r}\,\psi^*_{n\mathbf{k}}(\mathbf{r})\psi_{m\mathbf{q}}(\mathbf{r})+
\mathbf{d}_{nm}(\mathbf{k})\delta(\mathbf{k}-\mathbf{q})=\nonumber \\=&
-i\delta_{nm}\nabla_\mathbf{q}\delta(\mathbf{k}-\mathbf{q})+ 
\mathbf{d}_{nm}(\mathbf{k})\delta(\mathbf{k}-\mathbf{q}), 
\end{align}
where we introduced the notation 
\begin{align}
\mathbf{d}_{nm}(\mathbf{k})=\frac{i}{\Omega}\int_\Omega d^2\mathbf{r}\, 
u^*_{n\mathbf{k}}(\mathbf{r})\nabla_\mathbf{k}u_{m\mathbf{k}}(\mathbf{r}).
\end{align}
Note that $\mathbf{d}^*_{nm}(\mathbf{k})=\mathbf{d}_{mn}(\mathbf{k})$, 
and $\mathbf{d}_{nn}(\mathbf{k})\in \mathbb{R}$. This property can be proven with integration by parts
\begin{align}
\mathbf{d}^*_{nm}&(\mathbf{k})=-\frac{i}{\Omega}\int_\Omega d^2\mathbf{r}\, 
u_{n\mathbf{k}}(\mathbf{r})\nabla_\mathbf{k}u^*_{m\mathbf{k}}(\mathbf{r})= \nonumber \\ =&
-\frac{i}{\Omega}\nabla_\mathbf{k}\int_\Omega d^2\mathbf{r}\, 
u^*_{m\mathbf{k}}(\mathbf{r})u_{n\mathbf{k}}(\mathbf{r})+
\frac{i}{\Omega}\int_\Omega d^2\mathbf{r}\, 
u^*_{m\mathbf{k}}(\mathbf{r})\nabla_\mathbf{k}u_{n\mathbf{k}}(\mathbf{r}) \nonumber \\=&
-\frac{i}{\Omega} 
\nabla_\mathbf{k}\delta_{mn} +
\frac{i}{\Omega}\int_\Omega d^2\mathbf{r}\, 
u^*_{m\mathbf{k}}(\mathbf{r})\nabla_\mathbf{k}u_{n\mathbf{k}}(\mathbf{r})=\mathbf{d}_{mn}(\mathbf{k}).
\end{align} 
Substituting the result (\ref{eq:x_matrix_element}) into Eq.~(\ref{eq:h_int}), and using integration by parts 
in $\nabla_\mathbf{q}$-related term  we obtain 
\begin{align}
\mathcal{H}_\text{int}=&
i|e|\mathbf{E}(t)\sum_{n}\int_\text{BZ} d^2\mathbf{k}\,a^\dagger_{n\mathbf{k}}\nabla_\mathbf{k}a_{n\mathbf{k}}+ \nonumber \\+&
|e|\mathbf{E}(t)\sum_{n<m}\int_\text{BZ} d^2\mathbf{k}\,[\mathbf{d}_{nm}(\mathbf{k})
a^\dagger_{n\mathbf{k}}a_{m\mathbf{k}} + \mathbf{d}_{mn}(\mathbf{k})
a^\dagger_{m\mathbf{k}}a_{n\mathbf{k}}] + \nonumber \\+& 
|e|\mathbf{E}(t)\sum_{n}\int_\text{BZ} d^2\mathbf{k}\,\mathbf{d}_{nn}(\mathbf{k})
a^\dagger_{n\mathbf{k}}a_{n\mathbf{k}}.
\end{align}
Note that $\mathbf{d}_{mn}(\mathbf{k})$ for $n\neq m$ describes the coupling between $n$ and $m$ bands, 
while the term $\mathbf{d}_{nn}(\mathbf{k})$ is responsible for the intraband dynamics in $n$th bands.    
For the case of two bands $n,m=c,v$ this Hamiltonian gives the formula (\ref{eq:interaction}) in the main text of the study.

\section{Properties of the Berry connection in 2D semiconductors}
\label{app:berry_connection} 

We analyze the properties of the Berry connection 
\begin{align}
\label{eq:connection}
\mathbf{d}_{nn}(\mathbf{k})=\frac{i}{\Omega}\int_\Omega d^2\mathbf{r}\, 
u^*_{n\mathbf{k}}(\mathbf{r})\nabla_\mathbf{k}u_{n\mathbf{k}}(\mathbf{r})
\end{align}
at the $\mathbf{k}_0$ point within the $\mathbf{k}\cdot\mathbf{p}$ approach. To do so we 
present the Bloch function $u_{n\mathbf{k}}(\mathbf{r})$ at the point $\mathbf{k}$ as a superposition of the Bloch functions
$u_{n\mathbf{k}_0}(\mathbf{r})$ at the point $\mathbf{k}_0$, 
\begin{align}
\label{eq:decomposition}
u_{n\mathbf{k}}(\mathbf{r})=\sum_m C_{nm}(\mathbf{k}_0,\mathbf{k})u_{m\mathbf{k}_0}(\mathbf{r}).
\end{align}
Here we suppose that the set of functions $\{u_{m\mathbf{k}_0}(\mathbf{r})\}$ is known. Therefore, all the properties of the 
functions $u_{n\mathbf{k}}(\mathbf{r})$ are defined by the coefficients $C_{nm}(\mathbf{k}_0,\mathbf{k})$, which satisfy the normalization conditions 
\begin{align}
\delta_{nn'}=&\frac{1}{\Omega}\int_\Omega d^2\mathbf{r}\, u^*_{n\mathbf{k}}(\mathbf{r})u_{n'\mathbf{k}}(\mathbf{r})=
\nonumber \\=&
\sum_{mm'}
C^*_{nm}(\mathbf{k}_0,\mathbf{k})C_{n'm'}(\mathbf{k}_0,\mathbf{k})
\frac{1}{\Omega}\int_\Omega d^2\mathbf{r}\, u^*_{m\mathbf{k}_0}(\mathbf{r})u_{m'\mathbf{k}_0}(\mathbf{r})=
\nonumber \\=&
\sum_{mm'}
C^*_{nm}(\mathbf{k}_0,\mathbf{k})C_{n'm'}(\mathbf{k}_0,\mathbf{k})\delta_{mm'}=
\nonumber \\=&
\sum_{m}
C^*_{nm}(\mathbf{k}_0,\mathbf{k})C_{n'm}(\mathbf{k}_0,\mathbf{k}).
\end{align}
Substituting this decomposition (\ref{eq:decomposition}) into (\ref{eq:connection}) we obtain 
\begin{align}
\mathbf{d}_{nn}(\mathbf{k})=i\sum_m C^*_{nm}(\mathbf{k}_0,\mathbf{k})\nabla_\mathbf{k}C_{nm}(\mathbf{k}_0,\mathbf{k}).
\end{align}
To evaluate these coefficients we use the result from Ref.~[\onlinecite{Bir1974}] 
\begin{align}
\label{eq:bir}
u_{n\mathbf{k}_0+\boldsymbol{\kappa}}(\mathbf{r})=\frac{1}{A(\boldsymbol{\kappa})}\Big[u_{n,\mathbf{k}_0}(\mathbf{r})+\frac{\hbar}{m_0}
\sum_{\alpha, m\neq n}\frac{\kappa_\alpha p^\alpha_{mn}(\mathbf{k}_0)u_{m\mathbf{k}_0}}{E_n(\mathbf{k}_0)-E_m(\mathbf{k}_0)}\Big],
\end{align}
where $m_0$ is a free electron mass, $E_n(\mathbf{k}_0)$ is $n$th band energy at the point $\mathbf{k}_0$, and 
\begin{align}
p^\alpha_{mn}(\mathbf{k}_0)=\frac{1}{\Omega}\int_\Omega d^2\mathbf{r} u^*_{n\mathbf{k}_0}(\mathbf{r})\widehat{p}^\alpha u_{m\mathbf{k}_0}(\mathbf{r})
\end{align}
is ($nm$)th matrix element of the $\alpha(=x,y)$th component of the momentum operator $\widehat{\mathbf{p}}=-i\hbar (\partial_x, \partial_y)$, evaluated on the Bloch states with momentum $\mathbf{k}_0$.  
The coefficient $A(\boldsymbol{\kappa})$, where   
\begin{align}
A^2(\boldsymbol{\kappa})=1+\frac{\hbar^2}{m^2_0}
\sum_{\alpha,\beta, m\neq n}\frac{\kappa_\alpha\kappa_\beta  p^\alpha_{nm}(\mathbf{k}_0)p^\beta_{mn}(\mathbf{k}_0)}{[E_n(\mathbf{k}_0)-E_m(\mathbf{k}_0)]^2},
\end{align}
provides the normalization of the state $u_{n\mathbf{k}_0+\boldsymbol{\kappa}}(\mathbf{r})$ up to quadratic in $\boldsymbol{\kappa}=(\kappa_x,\kappa_y)$ parameter. 
Comparing (\ref{eq:bir}) with (\ref{eq:decomposition}) we obtain  
\begin{align}
C_{nn}(\mathbf{k}_0,\mathbf{k}_0+\boldsymbol{\kappa})=&\frac{1}{A(\boldsymbol{\kappa})}, \\
C_{nm}(\mathbf{k}_0,\mathbf{k}_0+\boldsymbol{\kappa})=&\frac{1}{A(\boldsymbol{\kappa})}\frac{\hbar}{m_0}
\sum_\alpha \frac{\kappa_\alpha p^\alpha_{mn}(\mathbf{k}_0)}{E_n(\mathbf{k}_0)-E_m(\mathbf{k}_0)},
\end{align}
for $n=m$ and $m\neq n$, respectively. Hence, we obtain 
\begin{align}
\big[\mathbf{d}_{nn}(\mathbf{k})\big]_\beta=\frac{i\hbar^2}{2m_0^2}
\sum_{\alpha, m\neq n}\kappa_\alpha
\frac{p^\alpha_{nm}(\mathbf{k}_0)p^\beta_{mn}(\mathbf{k}_0)-p^\beta_{nm}(\mathbf{k}_0)p^\alpha_{mn}(\mathbf{k}_0)}{[E_n(\mathbf{k}_0)-E_m(\mathbf{k}_0)]^2}
\end{align}
This expression defines the Berry curvatures $\boldsymbol{\Omega}_n(\mathbf{k})$ of the $n$th bands via the formula 
\begin{align}
\boldsymbol{\Omega}_n(\mathbf{k})\equiv&\nabla_\mathbf{k}\times \mathbf{d}_{nn}(\mathbf{k})=\mathbf{e}_z\Omega_n(\mathbf{k}_0)=\nonumber \\=&\mathbf{e}_z\frac{i\hbar^2}{m_0^2}
\sum_{m\neq n}\frac{p^x_{nm}(\mathbf{k}_0)p^y_{mn}(\mathbf{k}_0)-p^y_{nm}(\mathbf{k}_0)p^x_{mn}(\mathbf{k}_0)}{[E_n(\mathbf{k}_0)-E_m(\mathbf{k}_0)]^2}.  
\end{align}
The latter result coincides with the expressions presented in the literature \cite{Xiao2010,Chang2008}. Taking the latter result 
into account we write the Berry connection near $\mathbf{k}_0$ point 
\begin{align}
\mathbf{d}_{nn}(\mathbf{k})\approx \frac{\Omega_n(\mathbf{k_0})}{2}(-\kappa_y\mathbf{e}_x+\kappa_x\mathbf{e}_y). 
\end{align}
Here, $\mathbf{e}_x,\mathbf{e}_y,\mathbf{e}_z$ are the unit vectors in the $x,y$ and $z$ directions. 

We apply the formula obtained for the $\pm \mathbf{K}$ points in the transition metal dichalcogenide monolayer, that is, $\mathbf{k}_0\rightarrow \pm\mathbf{K}$. Motivated by the structure of the transition dipole moments at these points, we rewrite $\Omega_n(\mathbf{k}_0)$ in the form 
\begin{align}
\Omega_n(\mathbf{k}_0)=-\frac{\hbar^2}{4m_0^2}\sum_{m\neq n} 
\frac{|p^+_{nm}(\mathbf{k}_0)|^2 - |p^-_{nm}(\mathbf{k}_0)|^2}{[E_n(\mathbf{k}_0)-E_m(\mathbf{k}_0)]^2}, 
\end{align}
where $p^\pm_{nm}=p^x_{nm}\pm ip^y_{nm}$. The optical selection rules dictate $p^\pm_{nm}(\mathbf{K})=p^\mp_{nm}(-\mathbf{K})$. 
Therefore, $\Omega_n(\mathbf{K})=-\Omega_n(-\mathbf{K})$ and the Berry connection in $\tau=\pm1$ valleys reads
\begin{align}
\mathbf{d}_{nn}(\tau\mathbf{K}+\boldsymbol{\kappa})\approx \frac{\tau\Omega_n(\mathbf{K})}{2}(-\kappa_y\mathbf{e}_x+\kappa_x\mathbf{e}_y). 
\end{align}
The magnitude of the Berry curvature in the S-TMD monolayer is peaked only in the vicinity of the $\pm\mathbf{K}$ points and decays when moving away from them \cite{Cao2012,Liu2015}. Therefore, we approximate the Berry curvature as a constant, $\tau\Omega_n(\mathbf{K})$, within a circular domain of radius $\kappa_0$ centered at the point $\tau\mathbf{K}$, and assume it to be zero outside this domain. Moreover, we also use this approach for the Berry connection too. The size $\kappa_0$ of the corresponding 
``flux tube'' is a free parameter, which we equate to the size of the $1s$ exciton state in momentum space.  
Summarising, we obtain the following expression for the Berry connection at the $\tau\mathbf{K}$ point 
\begin{align}
\mathbf{d}_{nn}(\tau\mathbf{K}+\boldsymbol{\kappa})=\frac{\tau\Omega_n(\mathbf{K})}{2}\theta(\kappa_0-\kappa)(-\kappa_y\mathbf{e}_x+\kappa_x\mathbf{e}_y). 
\end{align}
Here, $\kappa=|\boldsymbol{\kappa}|=\sqrt{\kappa_x^2+\kappa_y^2}$, and $\theta(x)$ is a step function. 
Then, using the definition from Sec.~\ref{sec:sbe_2d} 
\begin{align}
\mathbf{D}^\tau(\mathbf{k})\equiv\mathbf{d}_{cc}(\tau\mathbf{K}+\mathbf{k})-\mathbf{d}_{vv}(\tau\mathbf{K}+\mathbf{k}),
\end{align}
and the derived expression for $\mathbf{d}_{nn}(\tau\mathbf{K}+\boldsymbol{\kappa})$, and applying it for $n=c,v$, 
we obtain 
\begin{align}
\mathbf{D}^\tau(\mathbf{k})=\tau\frac{\Omega_c(\mathbf{K})-\Omega_v(\mathbf{K})}{2}\theta(\kappa_0-k)(-k_y\mathbf{e}_x+k_x\mathbf{e}_y).
\end{align}
Comparing this result with the definition $\mathbf{D}^\tau (\mathbf{k}) = \tau D(k)(-k_y\mathbf{e}_x + k_x\mathbf{e}_y)$ 
from Sec.~\ref{sec:pump} we obtain 
\begin{align}
D(k)=\frac{\Omega_c(\mathbf{K})-\Omega_v(\mathbf{K})}{2}\theta(\kappa_0-k),
\end{align}
where $\kappa_0$ is the parameter which depends on the material properties of considered S-TMD monolayer.

\section{Alternative derivation of the Eq.~(\ref{eq:new_functions})}
\label{app:derivation}

We consider the secular equation (\ref{eq:matrix_secular_equation}), and 
where the coupling between $np$ and $nd$ states is neglected, i.e the states 
with $m=\pm 1$ and $m'=\pm 2$ angular momenta. In this case only $ns$ and $np$ states remain coupled, 
and hence each eigenfunction of the problem should be a superposition of the $s$ and $p$ states
\begin{align}
\label{eq:truncated_sum}
\widetilde{\psi}(\mathbf{k})=\sum_{n} a_n\psi_{n0}(\mathbf{k})+
\sum_{n} b_{n,\pm}\psi_{n,\pm1}(\mathbf{k}). 
\end{align} 
The general secular equation then reduces to the following set of the equations
\begin{align}
&Ea_n=(E_{ns}-\tau\hbar\omega_\text{p})a_n+\frac{ie\mathcal{E}_\text{p}}{2}
\sum_{n',\pm}[\lambda_{0,\pm1}]_{nn'}b_{n',\pm}, \\
&(E_{np}-\tau\hbar\omega_\text{p}\mp\hbar\omega_\text{p}-E)b_{n,\pm}=
\frac{ie\mathcal{E}_\text{p}}{2}\sum_{n'}[\lambda^\dagger_{0,\pm1}]_{nn'}a_{n'}.
\end{align} 
Evaluating $b_{n,\pm}$ from the second equation, and substituting it into the first equation, 
we obtain the equation for $a_n$ and spectrum $E$
\begin{align}
Ea_n=(E_{ns}-\tau\hbar\omega_\text{p})a_n-\frac{e^2\mathcal{E}^2_\text{p}}{4}
\sum_{n',n'',\pm}\frac{[\lambda_{0,\pm1}]_{nn'}[\lambda_{0,\pm1}^\dagger]_{n'n''}a_{n''}}
{E_{n'p}-\tau\hbar\omega_\text{p}\mp\hbar\omega_\text{p}-E}.
\end{align}
This equation is valid for any eigenfunction, represented by an infinite set of numbers $a_1,a_2, \dots b_{1,\pm}, b_{2,\pm} \dots$. 

Let us consider the particular case of $Ns$ state. In this case $a_n\rightarrow \delta_{nN}$, and $E\rightarrow \widetilde{E}_{Ns}$  
Then supposing that the energy of $Ns$ exciton is not significantly changed,
we make a substitution $\widetilde{E}_{Ns}\rightarrow E_{Ns}-\tau\hbar\omega_\text{p}+\Delta E_{Ns}$
and skip the term $\Delta E_{Ns}$ in the denominator of the sum.  
In this case, only one term in the sum gives the contribution and we obtain 
\begin{align}
\Delta E_{Ns}=-
\frac{e^2\mathcal{E}^2_\text{p}}{4}
\sum_{n',\pm}\frac{[\lambda_{0,\pm1}]_{Nn'}[\lambda^\dagger_{0,\pm1}]_{n'N}}
{E_{n'p}-E_{Ns}\mp \hbar\omega_\text{p}}.
\end{align}  
In this case the new $Ns$ state $\widetilde{\psi}_{N0}(\mathbf{k})$ has a form, 
where only the $N$-th component of $a_n=\delta_{nN}$ in the sum (\ref{eq:truncated_sum}) is nonzero. 
In this case we have   
\begin{align}
b_{n',\pm}=i\frac{e\mathcal{E}_\text{p}}{2}
\frac{[\lambda^\dagger_{0,\pm1}]_{n'N}}{E_{n'p}-E_{Ns}\mp\hbar\omega_\text{p}}.
\end{align} 
Therefore, the eigenfunction of the modified $Ns$ state is 
\begin{align}
\widetilde{\psi}_{N0}(\mathbf{k})=Z^{-1}_{N0}\Big[\psi_{N0}(\mathbf{k})+
i\frac{e\mathcal{E}_\text{p}}{2}\sum_{n',\pm} 
\frac{[\lambda^\dagger_{0,\pm1}]_{n'N}\psi_{n',\pm1}(\mathbf{k})}
{E_{n'p}-E_{Ns}\mp\hbar\omega_\text{p}}\Big], 
\end{align}
with 
\begin{align}
Z_{N0}^2=1+\frac{e^2\mathcal{E}^2_\text{p}}{4}\sum_{n',\pm} 
\frac{[\lambda_{0,\pm1}]_{Nn'}[\lambda^\dagger_{0,\pm1}]_{n'N}}{(E_{n'p}-E_{Ns}\mp\hbar\omega_\text{p})^2}.
\end{align}

\bibliography{aip_intraband}

\end{document}